\documentclass[iicol,pdflatex,sn-mathphys-num]{sn-jnl}% Math and Physical Sciences 
\usepackage{graphicx}%
\usepackage{multirow}%
\usepackage{amsmath,amssymb,amsfonts}%
\usepackage{amsthm}%
\usepackage{mathrsfs}%
\usepackage[title]{appendix}%
\usepackage{xcolor}%
\usepackage{textcomp}%
\usepackage{manyfoot}%
\usepackage{booktabs}%
\usepackage{algorithm}%
\usepackage{algorithmicx}%
\usepackage{algpseudocode}%
\usepackage{listings}%

\usepackage{booktabs}       % professional-quality tables
\usepackage{amsfonts,amsmath,amssymb, dutchcal}
\usepackage{nicefrac}       % compact symbols for 1/2, etc.
\usepackage{xcolor}         % colors
\usepackage[nolist]{acronym}
\usepackage{ragged2e}

\usepackage{nomencl}

\makenomenclature

\usepackage{etoolbox}
\renewcommand\nomgroup[1]{%
  \item[\bfseries
  \ifstrequal{#1}{A}{Conventions}{%%
  \ifstrequal{#1}{B}{Acoustics}{}}{%
  \ifstrequal{#1}{C}{Signal processing}{}}{
  \ifstrequal{#1}{D}{Special functions}{}}
]}
\usepackage{tikz}
\usepackage{pgfplots}
\usepgfplotslibrary{statistics}
\usetikzlibrary{calc}
\usetikzlibrary{shapes.geometric} % for shape=ellipse
\usepackage[normalem]{ulem}
\usepackage{array}
\usepackage{venndiagram}
\DeclareMathOperator*{\argmax}{arg\,max}

\algnewcommand\algorithmicforeach{\textbf{while}}
\algdef{S}[FOR]{While}[1]{\algorithmicforeach\ #1\ \algorithmicdo}

\def\sectionautorefname{Sec.}
\def\subsectionautorefname{Sec.}
\def\figureautorefname{Fig.}
\def\algorithmautorefname{Alg.}
\newcommand{\Autoref}[1]{%
  \begingroup%
  \renewcommand\equationautorefname{Equation}%
  \renewcommand\algorithmautorefname{Algorithm}%
  \renewcommand\figureautorefname{Figure}%
  \renewcommand\sectionautorefname{Section}%
  \renewcommand\subsectionautorefname{Section}%
  \autoref{#1}%
  \endgroup%
  }

\newcommand\utimes{\mathbin{\ooalign{$\cup$\cr%
   \hfil\raise0.42ex\hbox{$\scriptscriptstyle\times$}\hfil\cr}}}
\begin{acronym}
\acro{3GPP}{3rd Generation Partnership Project}
\acro{a-SLAM}[aSLAM]{Acoustic \aclu{SLAM}}
\acro{aSLAM}{Acoustic \aclu{SLAM}}
\acro{A-SNR}{A-weighted \acs{SNR}}
\acro{AAC}{Advanced Audio Coding\acroextra{. A lossy codec used for digital audio.}}
\acro{AAD}{Auditory Attention Detection}
\acro{AAS}{American Auditory Soc.}
\acro{AASP}{Audio and Acoustic Signal Processing}
\acro{ABC}{Analytical with or without Bias Compensation}
\acro{ABR}{Auditory-Brainstem Response}
\acro{ACAWD}{Archivable Core Actual-Word Database}
\acro{ACB}{Adaptive Codebook}
\acro{ACC}{Accuracy}
\acro{ACE}{Acoustic Characterization of Environments\acroextra{. A noisy reverberant speech corpus and IEEE challenge run by the SAP group at Imperial College}}
\acro{ACELP}{Algebraic Code-Excited Linear Prediction}
\acro{ACF}{Autocorrelation Function}
\acro{ACK}{Acknowledgement}
\acro{ACL}{Access Control List}
\acro{ACR}{Absolute Category Rating}
\acro{AD}{Audio Diarization}
\acro{ADC}{Analogue-to-Digital Converter}
\acro{ADM}{Adaptive Differential Microphone}
\acro{ADPCM}{Adaptive Differential Pulse Code Modulation}
\acro{ADSL}{Asymmetric Digital Subscriber Line}
\acro{AE}{Almost Everywhere}
\acro{AES}{Audio Engineering Society}
\acro{AES2}[AES]{Advanced Encryption Standard}
\acro{AGC}{Automatic Gain Control}
\acro{AH}{Amplitude Histogram}
\acro{AI}{Articulation Index}
\acro{AI2}[AI]{Artificial Intelligence}
\acro{AI3}[AI]{Audio Inpainting}
\acro{AIC}{Akaike Information Criterion}
\acro{AIFF}{Audio Interchange File Format}
\acro{AIR}{Acoustic Impulse Response}
\acro{AIR2}[AIR]{Aachen Impulse Response}
\acro{AIRD}{Aachen Impulse Response Database}
\acro{AIV}{Augmented Intensity Vectors}
\acro{ALC}{Automatic Level Control}
\acro{ALCons}{Articulation Loss of Consonants}
\acro{AM}{Amplitude Modulation}
\acro{AMDF}{Average Magnitude Difference Function\acroextra{. A function with similar properties to the cross- or autocorrelation but that requires no multiplication to evaluate.}}
\acro{AMR}{Adaptive Multi-Rate}
\acro{AMR-NB}{Adaptive Multi-Rate Narrow Band}
\acro{AMR-WB}{Adaptive Multi-Rate Wide Band}
\acro{AMS}{Amplitude Modulation Spectrogram}
\acro{ANC}{Adaptive Noise Canceller}
\acro{ANS}{Autocorrelation-Based Noise Subtraction}
\acro{ANSI}{American National Standards Institute}
\acro{ANU}{Australian National University}
\acro{APLAWD}{Archivable Priority List Actual-Word Database}
\acro{AR}{Autoregressive}
\acro{ARD}{Arbeitsgemeinschaft der \"{o}ffentlich-rechtlichen Rundfunkanstalten der Bundesrepublik Deutschland\acroextra{. `Consortium (``Working group'') of the public-law broadcasting institutions of the Federal Republic of Germany'}}
\acro{ARMA}{Autoregressive Moving Average}
\acro{AS}{Audio Segmentation}
\acro{AS2}[AS]{Almost Surely}
\acro{ASA}{Acoustic Scene Analysis}
\acro{ASIO}{Audio Stream Input/Output\acroextra{. A computer soundcard protocol with low latency developed by Steinberg.}}
\acro{ASK}{Amplitude Shift Keying}
\acro{ASLP}{Audio, Speech, and Language Processing}
\acro{ASM}{Acoustic Scene Mapping}
\acro{ASR}{Automatic Speech Recognition}
\acro{ASS}{Approximate Spectrum Substitution}
\acro{AST}{Acoustic Source Tracking}
\acro{AST2}[AST]{Asymmetric Sampling in Time}
\acro{ATF}{Acoustic Transfer Function\acroextra{. The Fourier Transform of the \acs{RIR}.}}
\acro{ATLM}{Acoustic Tokenization and Language Modelling}
\acro{ATR}{Advanced Telecommunications Research Institute International\acroextra{, Kyoto, Japan}}
\acro{AUC}{Area under the curve}
\acro{AURORA}{Aurora Experimental Framework for the Evaluation of the Performance of Speech Recognition Systems under Noisy Conditions}
\acro{AUV}{Autonomous Underwater Vehicle}
\acro{AV}{Audio-Visual}
\acro{AWGN}{Additive White Gaussian Noise}
\acro{AWS}{Approximate Waveform Substitution}
\acro{BASIE}{Bayesian Adaptive Speech Intelligibility Estimation}
\acro{BCE}{Blind Channel Estimation}
\acro{BCL}{Bekesy Comfortable Loudness}
\acro{BCR}{Block-Coordinate Relaxation}
\acro{BEM}{Boundary Element Method}
\acro{BER}{Bit Error Rate}
\acro{BIBO}{Bounded-Input Bounded-Output}
\acro{BIC}{Bayesian Information Criterion}
\acro{Bk}{Berksons\acroextra{. A unit for measuring intelligibility.}}
\acro{BK}[B\&K]{Br{\"u}el and Kj{\ae}r}
\acro{BLSTM}{Bidirectional \acs{LSTM}}
\acro{BM}{Blocking Matrix}
\acro{BO-SCPHD}{Bearing-only \acs{SC-PHD}}
\acro{BO-SLAM}{Bearing-only \acs{SLAM}}
\acro{BOT}{Bearing-only tracking}
\acro{BP}{Basis Pursuit}
\acro{BPCC}{Basis Pursuit with Clipping Constraints}
\acro{BPDN}{Basis Pursuit Denoising}
\acro{BPM}{Beats Per Minute}
\acro{BPSK}{Binary Phase Shift Keying}
\acro{BR}{Barrodale and Roberts' (algorithm)}
\acro{BRI}{Basic Rate Index}
\acro{BSD}{Bark Spectral Distortion}
\acro{BSI}{Blind System Identification}
\acro{BSIM}{Binaural Speech Intelligbility Model}
\acro{BSS}{Blind Source Separation}
\acro{BSTOI}{Binaural \acs{STOI}}
\acro{BW}{Bandwidth}
\acro{BZ}{Back-to-Zero}
\acro{C4DM}{Centre for Digital Music}
\acro{C50}[$C_\textrm{50}$]{Clarity Index}
\acro{C-GFB}{Combination Gas-fired Boiler}
\acro{CART}{Classification and Regression Tree}
\acro{CASA}{Computational Auditory Scene Analysis}
\acro{CBR}{Constant Bit Rate}
\acro{CCC}{Cross-Correlation Coefficient}
\acro{CCCC}{DARPA CSR Corpus Coordinating Committee}
\acro{CCD}{Charge-Coupled Device}
\acro{CCI}{Call Clarity Index}
\acro{CCITT}{Consultative Committee for International Telephony and Telegraphy}
\acro{CCM}{Contralateral Competing Message}
\acro{CCR}{Comparison Category Rating}
\acro{CDB}{Constant Directivity Beamformer}
\acro{CDMA}{Code Division Multiple Access}
\acro{CELP}{Code-excited Linear Prediction}
\acro{CFRC}{Coarse-To-Fine Region Contraction}
\acro{CHIEF}{Combined Helmholtz Integral Equation Formulation}
\acro{CIT}{Constrained Initial Taps}
\acro{CL}{Clipping Level}
\acro{CLEAR}{Centre for Law Enforcement Audio Research}
\acro{CLID}{Cluster Identification Test}
\acro{CLT}{Central Limit Theorem}
\acro{CMA}{Constant Modulus Algorithm}
\acro{CMASI}{Coherence Modulated Acoustic Speckle Interferometry}
\acro{CMB}{Cosmic Microwave Background}
\acro{CNC}{Consonant-Nucleus-Consonant}
\acro{CNG}{Comfort Noise Generation}
\acro{CNN}{Convolutional Neural Network}
\acrodefplural{CNN}[CNNs]{Convolutional Neural Networks}
\acro{CODEC}{Coder-Decoder}
\acro{CPE}{Customer Premises Equipment}
\acro{CPHD}{Cardinalized \acs{PHD}}
\acro{CPU}{Central Processing Unit}
\acro{CPMA}{Coprime Microphone Array}
\acro{CRACD}{Codec-robust Automatic Clipping Detector}
\acro{CRC}{Cyclic Redundancy Check}
\acro{CRNN}{Convolutional Recurrent Neural Network}
\acrodefplural{CRNN}[CRNNs]{Convolutional Recurrent Neural Networks}
\acro{CS}{Channel Shortening}
\acro{CS2}[CS]{Compressive Sensing}
\acro{CSP}{Communications and Signal Processing}
\acro{CSR-WSJ}{Continuous Speech Recognition Wall Street Journal Phase 1\acroextra{ database}}
\acro{CST}{Connected Speech Test\acroextra{ speech corpus}}
\acro{CT}{Conversation Test}
\acro{CUDA}{Compute Unified Device Architecture}
\acro{CTTN}{Comparative Tolerance to Noise}
\acro{CV}{Coefficient of Variation}
\acro{CVNN}{Complex-Valued Neural Networks}
\acrodefplural{CV}{Coefficients of Variation}
\acro{CV2}[CV]{Constant Velocity}
\acro{CVC}{Consonant-Vowel-Consonant}
\acro{CW}{Continuous Wave}
\acro{CWM}{Centre-Weighted Median}
\acro{CWMY}{Centre-Weighted Myriad}
\acro{CWT}{Continuous Wavelet Transform}
\acro{DAC}{Digital-to-Analogue Converter}
\acro{DAM}{Diagnostic Acceptability Measure}
\acro{DAQ}{Data Acquisition}
\acro{DARPA}{Defense Advanced Research Projects Agency\acroextra{ of the United States Dept. of Defense}}
\acro{DARPA-RMD}{\acs{DARPA} 1000-Word Resource Management Database\acroextra{ for Continuous Speech Recognition}}
\acro{DAW}{Digital Audio Workstation}
\acro{dB}{Decibel}
\acro{dBFS}{\acs{dB} Full Scale}
\acro{DBN}{Deep Belief Network}
\acro{DBSTOI}{Deterministic \acs{BSTOI}}
\acro{DC}{Direct Current}
\acro{DCME}{Digital Circuit Multiplexing Equipment}
\acro{DCR}{Degradation Category Rating}
\acro{DCT}{Discrete Cosine Transform}
\acro{DDR}{Direct-to-diffuse ratio}
\acro{DDR3}{Double Data Rate Type Three}
\acro{DECT}{Digital European Cordless Telecommunication}
\acro{DeLILAH}{Detection of Clipping using Least Squares Residuals and Iterated Logarithm Amplitude Histogram}
\acro{DENBE}{\acs{DRR} Estimation using a Null-Steered Beamformer}
\acro{DET}{Detection Error Trade-off}
\acro{Dev}{Development\acroextra{ dataset of the \acs{ACE} Challenge}}
\acro{DFT}{Discrete Fourier Transform}
\acro{DI}{Directivity Index}
\acro{DI-NN}{Dual-Input Neural Network}
\acro{DirectX}{\acroextra{A programming interface developed by Microsoft for handling tasks related to multimedia.}}
\acro{DIRHA}{Distant-speech Interaction for Robust Home Applications\acroextra{ multi-microphone multi-language acoustic speech corpus}}
\acro{DMA}{Differential Microphone Array}
\acro{DMT}{Discrete Multi-Tone}
\acro{DMV}{Dynamically Managed Voice\acroextra{ system}}
\acro{DNN}{Deep Neural Network}
\acro{DNR}{Dynamic Noise Reduction}
\acro{DoA}{Direction-of-Arrival}
\acrodefplural{DoA}[DoAs]{Directions-of-Arrival}
\acro{DOA}{Direction-of-Arrival}
\acrodefplural{DOA}[DOAs]{Directions-of-Arrival}
\acro{DP}{Dynamic Programming}
\acro{DPCM}{Differential Pulse Code Modulation}
\acro{DPD}{Direct-Path Dominance}
\acro{DPD-MUSIC}{Direct-Path Dominance Multiple Signal Classification}
\acro{DR}{Douglas-Rachford}
\acro{DR2}{Dynamic Range}
\acro{DRM}{Diagnostic Rhyme Test}
\acro{DRR}{Direct-to-Reverberant Ratio}
\acro{DRNN}{Deep Recurrent Neural Net}
\acro{DRT}{Diagnostic Rhyme Test}
\acro{DSB}{Delay-and-Sum Beamformer}
\acro{DSOBM}{Deterministic \acs{SOBM}}
\acro{DSP}{Digital Signal Processing}
\acro{DSPS}{Double Sides Periodic Substitution}
\acro{DSR}{Distributed Speech Recognition}
\acro{DSWS}{Double Sides Waveform Substitution}
\acro{DTMF}{Dual Tone Multi-Frequency}
\acro{DTX}{Discontinued Transmission}
\acro{DWT}{Discrete Wavelet Transform}
\acro{EARS}{Embodied Audition for RobotS}
\acro{EBF}{Eigen-beamformer}
\acro{EBU}{European Broadcasting Union}
\acro{EC}{Echo Canceller}
\acro{EDC}{Energy Decay Curve}
\acro{EDR}{Energy Decay Relief}
\acro{EDF}{Energy Decay Function}
\acro{EEG}{Electroencephalography}
\acro{EER}{Equal Error Rate}
\acro{EFICA}{Efficient Fast Independent Component Analysis}
\acro{EF-NN}{Early Fusion Neural Network}
\acro{EIR}{Equalized Impulse Response}
\acro{EKF}{Extended Kalman Filter}
\acro{EKF-SLAM}[EKF-SLAM]{\acs{EKF} \acs{SLAM}}
\acro{EKF-SLAM2}[EKF-SLAM]{Extended Kalman Filter \acs{SLAM}}
\acro{EL}{Echo Loss}
\acro{ELF}{Extremely Low Frequency}
\acro{ELRA}{European Languages Research Association}
\acro{EM}{Estimation-Maximization\acroextra{. An iterative technique to solve certain optimization problems.}}
\acro{EMIB}{Eigenmike Microphone Interface Box}
\acro{ENF}{Electrical Network Frequency}
\acro{EPSRC}{Engineering and Physical Sciences Research Council}
\acro{EQ}{Equalisation}
\acro{ERB}{Equivalent Rectangular Bandwidth}
\acro{ERP}{Ear Reference Point (cf. ITU-T Rec. P.64 1999)}
\acro{ESA}{Early Stage Assessment}
\acro{ESM}{Equivalent Source Method}
\acro{ESPRIT}{Estimation of Signal Parameters via Rotational Invariance Techniques}
\acro{ETAN}{Equivalent Tolerance to Additional Noise} 
\acro{ETSI}{European Telecommunications Standards Institute}
\acro{EURASIP}{European Association for Signal Processing}
\acro{EUSIPCO}{European Signal Processing Conference}
\acro{Eval}{Evaluation\acroextra{ dataset of the \acs{ACE} Challenge}}
\acro{F1}{F1 Score}
\acro{FAR}{False Alarm Rate}
\acro{FastSLAM}[FastSLAM]{FActored Solution To Simultaneous Localization and Mapping}
\acro{FastSLAM2}[FastSLAM]{FActored Solution To \acs{SLAM}} 
\acro{FAU}{Friedrich-Alexander-Universit{\"a}t}
\acro{FB}{Forward-Backward}
\acro{FB2}[FB]{Fullband}
\acro{FBF}{Fixed Beamformer}
\acro{FBSS}{Forward-Backward Spatial Smoothing}
\acro{FCC}{Federal Communications Commission}
\acro{FC-NN}{Fully Connected Neural Network}
\acro{FDM}{Frequency Division Multiplexing}
\acro{FDR}{False Discovery Rate}
\acro{FDR2}[FDR]{Free-Decay Region}
\acro{FEC}{Forward Error Correction}
\acro{FEM}{Finite Element Method}
\acro{FFI}{Norwegian Defence Research Establishment}
\acro{FFT}{Fast Fourier Transform}
\acroindefinite{FFT}{an}{a}
\acro{FIFO}{First-In First-Out}
\acro{FIR}{Finite Impulse Response\acroextra{. A filter whose output is a weighted sum of past input values and whose system function contains only zeros and no poles.}}
\acro{FISM}{Fast Image Source Method}
\acro{FISST}{Finite Set STatistics}
\acro{FLOM}{Fractional Lower-Order Moments}
\acro{FLOS}{Fractional Lower-Order Statistics}
\acro{FM}{Frequency Modulation}
\acro{FMM}{Fast Multipole Method}
\acro{FN}{False Negative}
\acro{FN2}{Nth Formant}
\acro{FNR}{False Negative Rate}
\acro{FORTRAN}{The IBM Mathematical Formula Translating System}
\acro{FoV}{Field of View}
\acro{FP}{False Positive}
\acro{FPGA}{Field Programmable Gate Array}
\acro{FPR}{False Positive Rate}
\acro{FPS}{Frames Per Second}
\acro{FRI}{Finite Rate of Innovation}
\acro{FSB}{Filter-and-Sum Beamformer}
\acro{FSK}{Frequency Shift Keying}
\acro{FT}{Flat-Top}
\acro{FWER}{Familywise Error Rate}
\acro{FWSSNR}{Frequency-Weighted Segmental \acs{SNR}}
\acro{FWSSRR}{Frequency-Weighted \acs{SSRR}}
\acro{G.711}{\acs{PCM} of Voice Frequencies}
\acro{GARCH}{Generalized Auto-regressive Conditional Heteroscedasticity}
\acro{GBW}{Gain Bandwidth Product}
\acro{GCC}{Generalized Cross-Correlation}
\acro{GCC-PHAT}{Generalized Cross-Correlation with Phase Transform\acroextra{ method of estimating \acs{TDoA}}}
\acro{GCF}{Global Coherence Field}
\acro{GGD}{Generalized Gaussian Distribution}
\acro{GGD2}[G$\Gamma$D]{Generalised Gamma Distribution}
\acro{GM}{Gaussian Mixture}
\acro{GM-PHD}{Gaussian Mixture \acs{PHD}}
\acro{GMCA}{Generalized Morphological Component Analysis}
\acro{GMM}{Gaussian Mixture Model\acroextra{. An approximation to an arbitrary probability density function that consists of a weighted sum of Gaussian distributions}}
\acro{GNN}{Graph Neural Network}
\acro{GNSS}{Global Navigation Satellite System}
\acro{GPRS}{General Packet Radio Services}
\acro{GPS}{Global Positioning System}
\acro{GPU}{Graphics Processing Unit}
\acro{GSC}{Generalized Sidelobe Canceller}
\acro{GSM}{Global System for Mobile Communications}
\acro{GSM-EFR}{\acs{GSM} Enhanced Full Rate Codec}
\acro{GSM-FR}{\acs{GSM} Full Rate Codec}
\acro{GSM-HR}{\acs{GSM} Half Rate Codec}
\acro{GUI}{Graphical User Interface}
\acro{HAAC}{High Amplitude Audio Capture}
\acro{HATS}{Head and Torso Simulator}
\acro{HERB}{Harmonicity-based dEReverBeration}
\acro{HFT}{Hands-Free Terminal}
\acro{HI}{Hearing-Impaired}
\acro{HIE}{Helmholtz Integral Equation}
\acro{HISAS}{High resolution Interferometric \ac{SAS}}
\acro{HINT}{Hearing-in-Noise Test}
\acro{HLT}{Human Language Technology}
\acro{HMM}{Hidden Markov Model}
\acro{HOS}{Higher-Order Statistics}
\acro{HPF}{High-Pass Filter}
\acro{HR}{Half Rate (\acs{GSM} Codec)}
\acro{HRI}{Human-Robot Interaction}
\acro{HRTF}{Head-related Transfer Function}
\acro{HSA}{Hearing, Speech, Audio\acroextra{ technology group at Fraunhofer IDMT}}
\acro{HSD}{Hybrid Steepest Descent}
\acro{HT}{Hannan-Thomson}
\acro{HTK}{Hidden Markov Model Tool Kit}
\acro{IBM}{Ideal Binary Mask}
\acro{IC}{Interference Canceller}
\acro{ICA}{Independent Component Analysis}
\acro{ICASSP}{Intl. Conf. on Acoustics, Speech and Signal Processing}
\acro{ID}{Identifier}
\acro{IDMT}{Institute for Digital Media Technology}
\acro{iDEN}{Integrated Digital Enhanced Network}
\acro{IEC}{International Electrotechnical Commission}
\acro{IEEE}{Institute of Electrical and Electronics Engineers}
\acro{IET}{Institute of Engineering and Technology}
\acro{IETF}{Internet Engineering Task Force}
\acro{IDFT}{Inverse \ac{DFT}}
\acro{IFFT}{Inverse Fast Fourier Transform}
\acro{IHC}{Inner Hair Cell}
\acro{IHT}{Iterative Hard Thresholding}
\acro{IID}{Independent and Identically Distributed}
\acro{IIR}{Infinite Impulse Response\acroextra{. A filter whose output is a weighted sum of both past input and past output values and whose system function contains both poles and zeros.}}
\acro{IL}{Iterated Logarithm\acroextra{, the logarithm of the logarithm}}
\acro{ILAH}{Iterated Logarithm Amplitude Histogram\acroextra{ clipping detection method}}
\acro{ILD}{Interaural Level Difference}
\acro{IMCRA}{Improved Minima Controlled Recursive Averaging\acroextra{. A technique for blindly estimating the spectrum of additive noise in a signal.}}
\acro{IMD}{Inter-Modulation Distortion}
\acro{IMSI}{International Mobile Subscriber Identity}
\acro{IMU}{Inertial Measurement Unit}
\acro{INMD}{In-service Non-intrusive Measurement Device}
\acro{INTERSPEECH}{Annual Conference of the \acs{ISCA}}
\acro{IO}{Infinitely Often}
\acro{IP}{Internet Protocol}
\acro{IPA}{International Phonetic Association}
\acro{IPD}{Interaural Phase Difference}
\acro{IRM}{Ideal Ratio Mask}
\acro{IPP}{Integrated Performance Primitives}
\acro{IRS}{Inverse repeated Sequence\acroextra{. A pseudo random sequence used for impulse response measurement.}}
\acro{IRS2}[IRS]{Intermediate Reference System}
\acro{IS}{Importance Sampling}
\acro{ISCA}{International Speech Communication Association}
\acro{ISDN}{Integrated Services Digital Network}
\acro{ISFT}{Inverse \acl{SFT}}
\acro{ISO}{Intl. Organization for Standardization}
\acro{ISP}{Intensity Spectral Profile}
\acro{IST}{Iterative Soft Thresholding}
\acro{ISTFT}{Inverse Short Time Fourier Transform}
\acro{ISVR}{Institute of Sound and Vibration Research\acroextra{, Southampton University, UK}}
\acro{ITD}{Interaural Time Difference}
\acro{ITF}{Interaural Transfer Function}
\acro{ITU}{International Telecommunication Union}
\acro{ITUR}[ITU-R]{International Telecommunication Union\acroextra{ Radiocommunication Sector}}
\acro{ITUT}[ITU-T]{International Telecommunication Union\acroextra{ Telecommunication Standardisation Sector}}
\acro{IUWT}{Isotropic Undecimated Wavelet (Starlet) Transform}
\acro{IWAENC}{Intl. Workshop on Acoustic Signal Enhancement}
\acro{IWAENC_PRE_2012}[IWAENC]{Intl. Workshop Acoustic Echo and Noise Control}
\acro{IWASE}{Intl. Workshop on Acoustic Signal Enhancement}
\acro{JADE}{Joint Approximate Diagonalization of Eigen-Matrices}
\acro{JPDA}{Joint Probabilistic Data Association}
\acro{JPEG}{Joint Photographic Experts Group}
\acro{KDE}{Kernel Density Estimate}
\acro{KF}{Kalman Filter}
\acro{KFD}{Kernel Fisher Discriminant\acroextra{ Analysis}}
\acro{KL}{Karhunen-Lo{\'{e}}ve}
\acro{KLT}{Karhunen-Lo{\'{e}}ve Transform}
\acro{KST}{Kolmogorov-Smirnov Test}
\acro{LAD}{Least Absolute Deviation}
\acro{LAN}{Local Area Network}
\acro{LARS}{Least Angle Regression}
\acro{LAT}[L$_{\textrm{AT}}$]{Equivalent Continuous Sound Level\acroextra{. Also called Leq}}
\acro{LBR}{Low Bitrate Redundancy}
\acro{LC}{Local Criterion}
\acro{LCMP}{Linearly Constrained Minimum Power}
\acro{LCMV}{Linearly Constrained Minimum Variance}
\acro{LCWM}{Linear Combination of Weighted Medians}
\acro{LDC}{Linguistic Data Consortium}
\acro{LEM}{Loudspeaker-Enclosure-Microphone System}
\acro{Leq}[L$_{\textrm{eq}}$]{Equivalent Continuous Sound Level\acroextra{. Also called LAT}}
\acro{LF}{Liljencrants-Fant\acroextra{. The developers of a glottal waveform model}}
\acro{LHS}{Left-Hand Side}
\acro{LID}{Language Identification}
\acro{LIDAR}{Light Detection and Ranging}
\acro{LILAH}{\acs{LSR2}-\acs{ILAH}\acroextra{ clipping detection method}}
\acro{LIME}{LInear Predictive Multi-input Equalization\acroextra{ algorithm}}
\acro{LiNoPS}{Lightweight Noise Protection System}
\acro{LLN}{Law of Large Numbers}
\acro{LLR}{Log-Likelihood Ratio}
\acro{LLS}{Logarithmic Least Squares}
\acro{LMA}{Least Mean Absolute}
\acro{LMS}{Least Mean Squares\acroextra{ adaptive filter}}
\acro{LNA}{Low Noise Amplifier}
\acro{LoS}{Line-of-Sight}
\acro{LOT}{Listening-Only Test}
\acro{LP}{Linear Parameter}
\acro{LP2}[LP]{Linear Predictive}
\acro{LP3}[LP]{Linear Prediction}
\acro{LPC}{Linear Predictive Coding\acroextra{. An autoregressive model of speech production.}}
\acro{LQO}{Listening Quality Objective}
\acro{LREC}{Conf. on Language Resources and Evaluation}
\acro{LS}{Least-Squares}
\acro{LSA}{Log Spectral Amplitude}
\acro{LSB}{Lower Side-Band}
\acro{LSB2}[LSB]{Least Significant Bit}
\acro{LSD}{Log Spectral Distortion}
\acro{LSP}{Line Spectrum Pairs}
\acro{LSR}{Late Stage Review}
\acro{LSR2}[LSR]{Least Squares Residuals\acroextra{ clipping detection method}}
\acro{LSRT}{Least Squares Residuals with Thresholding\acroextra{ clipping detection method}}
\acro{LSTM}{Long Short-Term Memory}
\acro{LTASS}{Long Term Average Speech Spectrum}
\acro{LTI}{Linear Time Invariant}
\acro{LTP}{Long Term Prediction}
\acro{LU}{Loudness Unit}
\acro{LUFS}{Loudness Units Full-Scale}
\acro{MA}{Moving Average}
\acro{MAC}{Multiply Accumulate Operation}
\acro{MAD}{Median Absolute Deviation}
\acro{MAE}{Mean Absolute Error}
\acro{MAP}{Maximum \emph{a posteriori}}
\acro{MARDY}{Multichannel Acoustic Reverberation Database at York}
\acro{MARS}{Multivariate Adaptive Regression Splines}
\acro{MBF}{Matched Filter Beamformer}
\acro{MC}{Monte Carlo}
\acro{MCA}{Morphological Component Analysis}
\acro{MCC}{Matthew's Correlation Coefficient}
\acro{MCCC}{Multichannel Cross-Correlation}
\acro{MCEQ}{MultiChannel EQualisation}
\acro{MCS}{Multidimensional Colouration Space}
\acro{MDCT}{Modified Discrete Cosine Transform}
\acro{MDL}{Minimum Description Length}
\acro{MDS}{Multidimensional Scaling}
\acro{Mel}{\acroextra{A non-uniform frequency scale corresponding to perceived frequency. It is approximately linear at low frequencies and logarithmic at high frequencies.}}
\acro{MELP}{Mixed Excitation Linear Prediction}
\acro{MFCC}{Mel-frequency Cepstral Coefficients}
\acro{MFS}{Method of Fundamental Solutions}
\acro{MFSK}{Multi-Frequency Shift Keying}
\acro{MHT}{Multi-Hypotheses Tracking}
\acro{MI}{Mutual Information}
\acro{MIMO}{Multiple-Input-Multiple-Output}
\acro{MINT}{Multiple-input/output INverse Theorem}
\acro{MIRS}{Motorola Integrated Radio System}
\acro{MIT}{Massachusetts Institute of Technology}
\acro{MIT-LCS}{Massacchusetts Institute of Technology Laboratory for Computer Science}
\acro{ML}{Maximum Likelihood}
\acro{MLD}{Masking Level Difference}
\acro{MLE}{Maximum Likelihood Estimation}
\acro{ML-SSL}{Maximum Likelihood Sound Source Localization}
\acro{ML-TDoA}{Maximum Likelihood Time Difference of Arrival}
\acro{MLMF}{Machine Learning with Multiple Features}
\acro{MLP}{Multi-layer Perceptron}
\acro{MLS}{Maximum Length Sequence\acroextra{ of pseudo random bits.}}
\acro{MMSE}{Minimum Mean Squared Error}
\acro{MMT}{Multiscale Median Transform}
\acro{MNRU}{Modulated Noise Reference Unit}
\acro{MOM}{Mean of Maximum}
\acro{MOS}{Mean Opinion Score}
\acro{MOS-LQO}{Mean Opinion Score - Listening Quality Objective}
\acro{MP}{Matching Pursuit}
\acro{MP3}{\acs{MPEG}-2 Audio Layer III}
\acro{MPEG}{Moving Picture Experts Group}
\acro{MRF}{Markov Random Field}
\acro{MRP}{Mouth Reference Point (cf. ITU-T Rec. P.64 1999)}
\acro{MRT}{Modified Rhyme Test}
\acro{MS}{Minimum Statistics}
\acro{MSB}{Most Significant Bit}
\acro{MSC}{Mean Square Coherence}
\acro{MSE}{Mean Square Error}
\acro{MSN}{Multiple Subscriber Number}
\acro{MSNR}{Maximum \acs{SNR}}
\acro{MTF}{Modulation Transfer Function}
\acro{MTM}{Modified Trimmed Mean}
\acro{MUSCLE}{MeasUred Single-CLustEr}
\acro{MUSHRA}{Multi-stimuli Test with Hidden Reference and Anchor}
\acro{MUSHRAR}{Multi-stimuli Test with Hidden Reference and Anchor for Reverberation}
\acro{MUSIC}{Multiple Signal Classification}
\acro{MVDR}{Minimum Variance Distortionless Response\acroextra{ beamformer}}
\acro{MWF}{Multi-channel Wiener Filter}
\acro{NAH}{Nearfield Acoustic Holography}
\acro{NB}{Narrowband}
\acro{NCM}{Normalized Coherence Metric}
\acro{NH}{Normal-Hearing}
\acro{NIRA}{Non-Intrusive Room Acoustics}
\acro{NISE}{Non-Intrusive \acs{SNR} estimation}
\acro{NISI}{Non-Intrusive Speech Intelligibility Estimation}
\acro{NISQ}{Non-Intrusive Speech Quality Estimation}
\acro{NIST}{National Institute of Standards and Technology}
\acro{NL}{Noise Level}
\acro{NLA}{Non-Linear Approximation}
\acro{NLMS}{Normalized Least Mean Squares\acroextra{ adaptive filter}}
\acro{NMCFLMS}{Normalized Multichannel Frequency Domain Least Mean Square}
\acro{NMF}{Non-negative Matrix Factorization}
\acro{NOISEX-92}{Database to Study the Effect of Additive Noise on Speech Recognition Systems}
\acro{NOIZEUS}{Noisy Speech Corpus for Evaluation of Speech Enhancement Algorithms}
\acro{NOSRMR}{Normalized Overall \acs{SRMR}}
\acro{NOS}{Number of Sources}
\acro{NOSRMR}{Normalized Overall \acs{SRMR}}
\acro{NPM}{Normalized Projection Misalignment}
\acro{NPV}{Negative Predictive Value}
\acro{NR}{Noise Reduction}
\acro{NS}{Noise Suppression}
\acro{NSRMR}{Normalised \acs{SRMR}}
\acro{NSRR}{Normalized Signal-to-Reverberation Ratio}
\acro{NSRMR}{Normalised \acs{SRMR}}
\acro{NSV}{Negative-Side Variance}
\acro{NTP}{Network Time Protocol}
\acro{NV}{Noise-Vocoding}
\acro{OBL}{Octave Band Level}
\acro{ODF}{Overdrive Factor}
\acro{Ofcom}{Office of Communications\acroextra{, the independent regulator and competition authority for the UK communications industries}}
\acro{OFDM}{Orthogonal Frequency Division Multiplexing}
\acro{OHC}{Outer Hair Cell}
\acro{OIM}{Objective Intelligibility Measure}
\acro{OLA}{Overlap-add}
\acro{OM-LSA}{Optimally Modified Log-Spectral Amplitude{ Estimator}}
\acro{OMP}{Orthogonal Matching Pursuit}
\acro{OSI}{Open Systems Interconnection}
\acro{OSPA}{Optimal Subpattern Assignment}
\acro{OSRMR}{Overall \acs{SRMR}}
\acro{PAB-SRMR}{Per acoustic band \acs{SRMR}}
\acro{PALM}{Passive Acoustic Localization and Mapping}
\acro{PAMS}{Perceptual Analysis Measurement System}
\acro{PARCOR}{Partial Correlation Coefficients}
\acro{PB}{Phonetically Balanced}
\acro{PBF}{Positive Boolean Function}
\acro{PCA}{Principal Components Analysis}
\acro{PCM}{Pulse-Code Modulation}
\acro{PDA}{Personal Digital Assistant}
\acro{PDA2}[PDA]{Probabilistic Data Association}
\acro{PDE}{Partial Differential Equation}
\acro{pdf}{Probability Density Function}
\acro{PDF}{Probability Density Function}
\acro{PE}{Parameter Estimation}
\acro{PEASS}{Perceptual Evaluation for Audio Source Separation}
\acro{PEFAC}{Pitch Estimation Filter with Amplitude Compression}
\acro{PESQ}{Perceptual Evaluation of Speech Quality}
\acro{PF}{Psychometric Function}
\acro{pgfl}[p.g.fl.]{Probability Generating Functional}
\acro{PHAT}{Phase Transform}
\acro{PHD}{Probability Hypothesis Density}
\acro{PIP}{Peak-Image Pairing}
\acro{PIV}{Pseudo-Intensity Vector}
\acro{PL}{Pseudo-likelihood}
\acro{PLC}{Packet Loss Concealment}
\acro{PLL}{Phase Locked Loop}
\acro{PLP}{Perceptual Linear Prediction}
\acro{PLR}{Perceived Level of Reverberation}
\acro{PM}{Phase Modulation}
\acro{PMF}{Probability Mass Function}
\acro{PMOS}{Predicted Mean Opinion Score}
\acro{PNN}{Parameterized Neural Network}
\acro{POLQA}{Perceptual Objective Listening Quality Analysis}
\acro{POTS}{Plain Old Telephone Service}
\acro{PPP}{Poisson Point Process}
\acro{PPS}{Pulse-Per-Second}
\acro{PPV}{Positive Predictive Value}
\acro{PReLU}{Parametric Rectified Linear Unit}
\acro{PRLM}{Phoneme Recognition and Language Modelling}
\acro{PRP}{Pair-wise Relative Phase-ratio}
\acro{PSD}{Power Spectral Density}
\acro{PSK}{Phase Shift Keying}
\acro{PSNR}{Peak Signal-to-Noise Ratio}
\acro{PSOLA}{Pitch Synchronous Overlap Add\acroextra{. A method of scaling a signal in time and pitch independently.}}
\acro{PSQM}{Perceptual Speech Quality Measurement}
\acro{PSSL}{Positional Sound Source Localization}
\acro{PSTN}{Public Switched Telephone Network}
\acro{PWD}{Plane-Wave Decomposition}
\acro{PTA}{Pure-Tone Audiology}
\acro{QAM}{Quadrature Amplitude Modulation}
\acro{QILAH}{Quadrisected Iterated Logarithm Amplitude Histogram\acroextra{ clipping detection method}}
\acro{QMF}{Quadrature Mirror Filter}
\acro{QoE}{Quality-of-Experience}
\acro{QoS}{Quality-of-Service}
\acro{QPSK}{Quadrature Phase Shift Keying}
\acro{RADAR}{RAdio Detection And Ranging}
\acro{RASTA}{Relative Spectral}
\acro{RASTA-PLP}{Relative Spectral Perceptual Linear Prediction}
\acro{RASTI}{Room Acoustics Speech Transmission Index\acroextra{ (superseded by STIPA)}}
\acro{RBM}{Restricted Boltzmann Machine}
\acro{RBFN}{Radial Basis Function Network}
\acro{RB-PHD}{Rao-Blackwellised \acs{PHD}}
\acro{RBPF}{Rao-Blackwellised Particle Filter}
\acro{RC}{Relative Criterion}
\acro{RDT}[$R_\textrm{DT}$]{Reverberation Decay Tail}
\acro{RDTF}{Relative Direct Transfer Function}
\acro{ReLU}{Rectified Linear Unit}
\acro{RF}{Radio Frequency}
\acro{RFI}{Radio Frequency Interference}
\acro{RFS}{Random Finite Set}
\acro{RHS}{Right-Hand Side}
\acro{RIP}{Restricted Isometry Property}
\acro{RIR}{Room Impulse Response}
\acro{RLS}{Recursive Least Squares\acroextra{ adaptive filter}}
\acro{RLSD}{Relative Log Spectral Distortion}
\acro{RMCLS}{Relaxed Multichannel Least Squares}
\acro{RMCLS-CIT}{Relaxed MultiChannel Least-Squares with Constrained Initial Taps}
\acro{RMS}{Root Mean Square}
\acro{RMSE}{Root Mean Square Error}
\acro{RNN}{Recurrent Neural Network}
\acro{ROC}{Receiver Operating Characteristic}
\acro{ROHC}{Robust Header Compression}
\acro{RP}{Received Pronunciation}
\acro{RPE}{Regular Pulse Excitation}
\acro{RRTF}{Relative Real-Time Factor}
\acro{RS}{Reverberation Suppression}
\acro{RSM}{Reflector Source Method}
\acro{RSS}{Received Signal Strength}
\acro{RSV}{Room Spectral Variance}
\acro{RT}{Reverberation Time}
\acro{RTAN}{Robustness to Additional Noise}
\acro{RTF}{Real-Time Factor}
\acro{RTF2}[RTF]{Room Transfer Function}
\acro{RTF3}[RTF]{Relative Transfer Function}
\acro{RV}{Random Variable}
\acro{RVP}{Recursive Vector Projection}
\acro{RWTH}{Rheinisch-Westf\"{a}lische Technische Hochschule}
\acro{S50}[$S_{50}$]{Intelligibility Function Gradient at the \ac{SRT}}
\acro{SA}{Spectral Amplitude}
\acro{SAA}{Synthetic Aperture Audio}
\acro{SAP}{Speech And Audio Processing}
\acro{SAR}{Speech-to-Artifact Ratio}
\acro{SAR2}[SAR]{Speaker Alternation Rate}
\acro{SAR3}[SAR]{Synthetic Aperture \ac{RADAR}}
\acro{SAS}{Synthetic Aperture \ac{SONAR}}
\acro{SB}{Subband}
\acro{SC-PHD}{Single Cluster \acs{PHD}}
\acro{SCAF}{Single Channel Adaptive Filter}
\acro{SCB}{Stochastic Codebook}
\acro{SCNR}{Single-Channel Noise Reduction}
\acro{SCOT}{Smoothed Coherence Transform}
\acro{SCNR}{Single-Channel Noise Reduction}
\acro{SC-PHD}{Single Cluster \acs{PHD}}
\acro{SCR}{Signal-to-Competition Ratio}
\acro{SCRIBE}{Spoken Corpus of British English}
\acro{SCT}{Speech Corruption Toolkit}
\acro{SCT2}{Short Conversation Test}
\acro{SD}{Semantic Differential}
\acro{SDB}{Superdirective Beamformer}
\acro{SDD}{Spectral Decay Distributions}
\acro{SDDMSB}{\acs{SDD} with Mel-spaced frequency bands}
\acro{SDDSA}{\acs{SDD} with Mel-spaced frequency bands and selective averaging}
\acro{SDDSA-G}{\acs{SDDSA} with Gerkmann noise estimator}
\acro{SDDSA-H}{\acs{SDDSA} with Hendriks noise estimator}
\acro{SDR}{Software Defined Radio}
\acro{SDR2}{Speech}
\acro{SDRAM}{Synchronous Dynamic Random Access Memory}
\acro{SDT}{Speech Description Taxonomy}
\acro{SEDF}{Subband \ac{EDF}}
\acro{SELD}{Sound Event Localization and Detection}
\acro{SEMG}{Surface Electromyography}
\acro{SFDR}{Spurious Free Dynamic Range}
\acro{SFM}{Single Feature with Mapping}
\acro{SFT}{Spherical Fourier Transform}
\acro{SH}{Spherical Harmonic}
\acro{SHD}{Spectral Harmonic Decomposition}
\acro{SHD2}[SHD]{Spherical Harmonic Domain}
\acro{SIE}{System Identification Error}
\acro{SII}{Speech Intelligibility Index}
\acro{SIImod}{Speech Intelligibility Index in the modulation domain}
\acro{SIM}{Subscriber Identity Module}
\acro{SIMO}{Single-Input-Multiple-Output}
\acro{SINAD}{Signal-to-Noise and Distortion Ratio}
\acro{SIP}{Session Initiation Protocol}
\acro{SIR}{Signal-to-Interference Ratio}
\acro{SIR2}[SIR]{Sequential Importance Resampling}
\acro{SIREAC}{Simulation of REal Acoustics\acroextra{ Software Tool}}
\acro{SIS}{Sequential Importance Sampling}
\acro{SL}{Speech Level}
\acro{SLAM}{Simultaneous Localization and Mapping}
\acro{SLO}{Spatial Observability Function}
\acro{SLF}{Spatial Likelihood Function}
\acro{SLLN}{Strong Law of Large Numbers}
\acro{SLM}{Sound Level Meter}
\acro{SMA}{Spherical Microphone Array}
\acro{SMARD}{Single- and Multichannel Audio Recordings Database}
\acro{SMERSH}{Spatiotemporal Averaging Method for Enhancement of Reverberant Speech}
\acro{SMIR}{Spherical Microphone array Impulse Response}
\acro{SMPTE}{Society of Motion Picture and Television Engineers}
\acro{SMS}{Short Message Service}
\acro{SNR}{Signal-to-Noise Ratio}
\acro{SNR2}[SNR]{Speech-to-Noise Ratio}
\acro{SNT}{Subspace Noise Tracking\acroextra{ algorithm}}
\acro{SOBM}{STOI-optimal Binary Mask}
\acro{SOF}{Spatial Observability Function}
\acro{SONAR}{SOund Navigation And Ranging}
\acro{SOLA}{Synchronous Overlap Add\acroextra{. A method of scaling a signal in time and pitch independently.}}
\acro{SPC}{Specificity}
\acro{SPEECON}{Speech Databases for Consumer Devices}
\acro{SPHERE}{NIST SPeech Header REsources\acroextra{ software with embedded Shorten Compression}}
\acro{SPIN}{Speech Perception In Noise}
\acro{SPL}{Sound Pressure Level}
\acro{SPP}{Speech Presence Probability}
\acro{SPQA}{Speech Quality Assurance Package}
\acro{SQNR}{Signal-to-Quantization Noise Ratio}
\acro{SR}{Sparse Representation}
\acro{SR2}[SR]{Spectral Rotation}
\acro{SRA}{Statistical Room Acoustics}
\acro{SRC}{Stochastic Region Contraction}
\acro{SRI}{SRI International\acroextra{. Formerly Standford Research Institute}}
\acro{SRMR}{Speech-to-Reverberation Modulation Energy Ratio}
\acro{SRK}{Steered Response Kurtosis}
\acro{SRP}{Steered Response Power}
\acro{SRP-PHAT}{Steered Response Power with Phase Transform}
\acro{SRP-TDE}{Steered Response Power with Time Delay Estimation}
\acro{SRR}{Signal-to-Reverberation Ratio}
\acro{SRT}{Speech Reception Threshold\acroextra{ (also known as Speech Recognition Threshold)}}
\acro{SS}{Spectral Subtraction}
\acro{SSB}{Single Side-Band}
\acro{SSI}{Synthetic Sentence Identification}
\acro{SSL}{Sound Source Localization}
\acro{SSN2}[SSN]{Simultaneous Switching Noise}
\acro{SSN}{Speech-Shaped Noise}
\acro{SSNR}{Segmental \acs{SNR}}
\acro{SSOBM}{Stochastic \acs{SOBM}}
\acro{SSRR}{Segmental Signal-to-Reverberation Ratio}
\acro{SSW}{Staggered Spondaic Word}
\acro{STFT}{Short Time Fourier Transform}
\acro{STI}{Speech Transmission Index}
\acro{STIPA}{Speech Transmission Index for Public Address Systems}
\acro{STITEL}{Speech Transmission Index for Telecommunication Systems}
\acro{STMI}{Spectro-Temporal Modulation Index}
\acro{STNR}{\acs{NIST}'s Speech-to-Noise Ratio\acroextra{ Estimation Algorithm}}
\acro{STOI}{Short-Time Objective Intelligibility\acroextra{ measure}}
\acro{STQ}{Speech Processing, Transmission and Quality Aspects}
\acro{STSA}{Short Time Spectral Analysis}
\acro{STSA1}[STSA]{Short Time Spectral Amplitude}
\acro{SUS}{Semantically Unpredictable Sentences}
\acro{SVD}{Singular Value Decomposition}
\acro{SVM}{Support Vector Machine}
\acro{SWSOBM}{Stochastic \acs{WSOBM}}
\acro{T20}[$T_\textrm{20}$]{Reverberation Time\acroextra{ to decay by $20$ dB}}
\acro{T30}[$T_\textrm{30}$]{Reverberation Time\acroextra{ to decay by $30$ dB}}
\acro{T60}[$T_\textrm{60}$]{Reverberation Time\acroextra{ to decay by $60$ dB}}
\acro{TBD}{Track Before Detect}
\acro{TBM}{Target Binary Mask}
\acro{TDE}{Time Delay Estimation}
\acro{TDHS}{Time Domain Harmonic Scaling\acroextra{. A method of scaling a signal in time and pitch independently.}}
\acrodefplural{TDOA}[TDOAs]{Time-Differences-of-Arrival}
\acro{TDOA}{Time-Difference-of-Arrival}
\acrodefplural{TDOA}[TDOAs]{Time-Differences-of-Arrival}
\acro{TDT}{Tone Decay Test}
\acro{TF}{Time-Frequency}
\acro{TFS}{Temporal Fine Structure}
\acro{TFGM}{Time-Frequency Gain Modification\acroextra{. An approach to signal enhancement in which a signal is multiplied by a gain function in the time-frequency domain.}}
\acro{THD}{Total Harmonic Distortion}
\acro{TI}{Texas Instruments, Inc.}
\acro{TIMIT}{\acs{TI}-\acs{MIT} speech corpus}
\acro{TIPHON}{Telecommunication and Internet Protocol Harmonization Over Networks}
\acro{TLS}{Total Least-Squares}
\acro{TN}{True Negative}
\acro{TNR}{True Negative Rate}
\acro{TOA}{Time-of-Arrival}
\acrodefplural{TOA}[TOAs]{Times-of-Arrival}
\acro{TOF}{Time-of-Flight}
\acro{TOSQA}{Telekom Objective Speech Quality Assessmentt}
\acro{TP}{True Positive}
\acro{TP2}[TP]{Trivial Pursuit}
\acro{TPCC}{Trivial Pursuit with Clipping Constraints}
\acro{TPR}{True Positive Rate}
\acro{TSE}{Taylor Series Expansion}
\acro{TVAR}{Time-varying Autoregression}
\acro{UAV}{Unmanned Aerial Vehicle}
\acro{UDP}{User Datagram Protocol}
\acro{UFRJ}{Federal University of Rio de Janeiro}
\acro{UHF}{Ultra High Frequency}
\acro{UKF}{Unscented Kalman Filter}
\acro{ULA}{Uniform Linear Array}
\acro{ULF}{Ultra Low Frequency}
\acro{UMTS}{Universal Mobile Telecommunications Service}
\acro{US}{United States}
\acro{UTBM}{Universal Target Binary Mask}
\acro{VAD}{Voice Activity Detector}
\acro{VBR}{Variable Bit-Rate}
\acro{VCV}{Vowel-Consonant-Vowel}
\acro{VRD}{Variance of Decay-rates}
\acro{VGC}{Voice Grade Channel}
\acro{vMF}{von Mises-Fisher}
\acro{VoIP}{Voice Over Internet Protocol}
\acro{VRT}{Vlaamse Radio- en Televisieomroeporganisatie\acroextra{. (Flemish Radio and Television Broadcasting Organization)}}
\acro{VSELP}{Vector Sum-excited Linear Prediction}
\acro{V-SRP}{Volumetric \acl{SRP}}
\acro{VST}{Virtual Studio Technology\acroextra{. An interface standard developed by Steinberg for adding plugins to an audio editor.}}
\acro{WADA}{Waveform Amplitude Distribution Analysis}
\acro{WASN}{Wireless Acoustic Sensor Network}
\acrodefplural{WASN}[WASNs]{Wireless Acoustic Sensor Networks}
\acro{WASPAA}{Workshop on Applications of Signal Processing to Audio and Acoustics}
\acro{WAV}{Waveform Audio File Format}
\acro{WAVE}{Waveform Audio File Format}
\acro{WB}{Wideband}
\acro{WER}{Word Error Rate}
\acro{WGN}{White Gaussian Noise}
\acro{WLAN}{Wireless \acs{LAN}}
\acro{WLLN}{Weak Law of Large Numbers}
\acro{WM}{Working Memory}
\acro{WMA}{Windows Media Audio}
\acro{WNG}{White Noise Gain}
\acro{WSS}{Weighted Spectral Slope}
\acro{WSOBM}{Weighted \acs{SOBM}}
\acro{WSTOI}{Weighted \acs{STOI}}
\acro{XSRP}{eXtensible \acs{SRP}}
\acro{ZOS}{Zero-Order Statistics}
\end{acronym}

\begin{document}

\title[Steered Response
Power for Sound Source
Localization: \\ A Tutorial Review]{Steered Response
Power for Sound Source
Localization: \\ A Tutorial Review}


\author*[1]{\fnm{Eric} \sur{Grinstein}}\email{e.grinstein@imperial.ac.uk}

\author[2]{\fnm{Elisa} \sur{Tengan}}%\email{elisa.tengan@esat.kuleuven.be}

\author[2]{\fnm{Bilgesu} \sur{Çakmak}}%\email{bilgesu.cakmak@esat.kuleuven.be}

\author[2]{\fnm{Thomas} \sur{Dietzen}}%\email{thomas.dietzen@esat.kuleuven.be}
\author[3]{\fnm{Leonardo} \sur{Nunes}}%\email{lnunes@microsoft.com}
\author[2]{\fnm{Toon} \sur{van Waterschoot}}%\email{toon.vanwaterschoot@esat.kuleuven.be}
\author[1]{\fnm{Mike} \sur{Brookes}}%\email{mike.brookes@imperial.ac.uk}
\author[1]{\fnm{Patrick} \sur{A. Naylor}}%\email{p.naylor@imperial.ac.uk}

\affil*[1]{\orgdiv{Department of Electrical and Electronic Engineering}, \orgname{Imperial College London}, \orgaddress{\country{U.K.}}}

\affil[2]{\orgdiv{Department of Electrical Engineering (ESAT), STADIUS Center for
Dynamical Systems, Signal Processing, and Data Analytics}, \orgname{KU Leuven}, \orgaddress{\country{Belgium}}}

\affil[3]{\orgname{Microsoft Research}, \orgaddress{\city{Rio de Janeiro}, \country{Brazil}}}


\abstract{In the last three decades, the \acf{SRP} method has been widely used for the task of \acf{SSL}, due to its satisfactory localization performance on moderately reverberant and noisy scenarios. Many works have analyzed and extended the original \ac{SRP} method to reduce its computational cost, to allow it to locate multiple sources, or to improve its performance in adverse environments. In this work, we review over 200 papers on the \ac{SRP} method and its variants, with emphasis on the \ac{SRP-PHAT} method. We also present eXtensible-SRP, or X-SRP, a generalized and modularized version of the SRP algorithm which allows the reviewed extensions to be implemented. We provide a Python implementation of the algorithm which includes selected extensions from the literature.}


\maketitle

% General notation
\nomenclature[A, a]{$r$}{Scalars are represented as italic letters}
\nomenclature[A, b]{$\mathbf{r}$}{Vectors are represented in boldface}
\nomenclature[A, c]{$\hat{r},\tilde{r}$}{Estimate or Approximation of variable $r$}
\nomenclature[A, cc]{$\bar{r}$}{Frequency-domain quantities are marked with a top bar}
\nomenclature[A, e]{$\mathbf{r}[k]$}{k-th element of vector $\mathbf{r}$}
\nomenclature[A, f]{$\mathbf{R}$}{Matrices are represented using uppercase bold letters}
\nomenclature[A, g]{$\mathcal{R},\mathcal{R}(t)$}{Sets and custom functions are represented using uppercase caligraphic letters}

% Acoustics
\nomenclature[B, a]{$c$}{Speed of sound}
\nomenclature[B, b]{$N$}{Number of active sources}
\nomenclature[B, c]{$M$}{Microphone array size}
\nomenclature[B, ca]{$P$}{Number of microphone pairs}

\nomenclature[B, f]{$\mathbf{u}_n$}{Cartesian coordinates of sound source $n$. In the single source problem, $\mathbf{u}$ is used. }
\nomenclature[B, g]{$\mathbf{v}_m$}{Cartesian coordinates of microphone $m$}
\nomenclature[B, h]{$\mathcal{G}$}{Set of candidate source positions forming the search grid}

\nomenclature[B, i]{$a_{m}(\mathbf{u})$}{Propagation attenuation from a source at $\mathbf{u}$ to microphone $m$.}
\nomenclature[B, j]{$\tau_{m}(\mathbf{u})$}{Propagation delay from a source at $\mathbf{u}$ to microphone $m$.}
\nomenclature[B, k]{$\tau_{lm}(\mathbf{u})$}{TDOA between microphones $l$ and $m$ to a source at $\mathbf{u}$}

\nomenclature[B, i]{$h_m(t; \mathbf{u})$}{Room Impulse Response between a source at $\mathbf{u}$ and microphone $m$.}

% Signals
\nomenclature[C, a]{$f_s$}{Sampling rate}
\nomenclature[C, aa]{$L$}{Signal frame size}
\nomenclature[C, b]{$s_n(t)$}{Sample of signal emitted by source $n$ at time $t$}
\nomenclature[C, c]{$x_m(t),\mathbf{x}_m(t)$}{Received sample or frame at time $t$ for microphone $m$}
\nomenclature[C, d]{$\bar{x}_m(t, f)$}{Received signal sample at time-frequency bin $t, f$ for microphone $m$}
\nomenclature[C, e]{$\bar{\mathbcal{x}}_m(t)$}{Received frequency-domain frame at time $t$ for microphone $m$}
\nomenclature[C, f]{$\mathbf{g}_t(t),\bar{\mathbf{g}}_f(t)$}{Time or frequency GCC-PHAT vector at time $t$}
\nomenclature[C, g]{$\mathcal{X}(t)$}{The set of microphone signal frames at time $t$}

% Functions
\nomenclature[D, a]{$\text{DFT}(\mathbf{x}),\text{IDFT}(\bar{\mathbf{x}})$}{Forward and inverse Discrete Fourier Transform operation}
\nomenclature[D, aa]{$\text{STFT}(\mathbf{x}),\text{ISTFT}(\bar{\mathbf{x}})$}{Forward and inverse Short-time Fourier Transform operation}
\nomenclature[D, ab]{$\text{CC}(\tau ; \mathbf{x}_l, \mathbf{x}_m)$}{Temporal cross-correlation function of signals $\mathbf{x}_l$ and $\mathbf{x}_m$ at time lag $\tau$}
\nomenclature[D, b]{$\text{GCC}(\tau ; \bar{\mathbf{x}}_l, \bar{\mathbf{x}}_m)$}{GCC-PHAT function between signals $\bar{\mathbf{x}}_l$ and $\bar{\mathbf{x}}_m$ at time lag $\tau$}
\nomenclature[D, c]{$\text{SRP}(\mathbf{u} \, ; \,\, \mathcal{X})$}{Temporal SRP-PHAT function for set of signals $\mathcal{X}$ evaluated at candidate location $\mathbf{u}$}

\printnomenclature

\section{Introduction} \label{sec:introduction}

% Intro: SSL, other methods
\acf{SSL} is the task of estimating the position of one or more active acoustic sources using one or more microphone arrays. Applications for \ac{SSL} include event detection \cite{Li2018c, Li2012a, Lopez-Morillas2016a}, camera steering \cite{Marti2011b}, and sound source separation \cite{Do2011a, Dam2016a, Wu2021a} among many others. In the last decades, many classical signal processing-based methods were developed for \ac{SSL}, including \ac{MUSIC} \cite{Schmidt1986c}, \ac{ESPRIT} \cite{Roy1989b}, \ac{TDOA}-based \cite{Brandstein1997e, Gustafsson2003a}, \ac{ML}-based \cite{So2011c} and \acf{SRP} \cite{Omologo1994a, DiBiase2000a}, which is the focus of this review. Alternatively to signal processing-based methods, significant research interest has also been devoted to machine learning-based localization methods \cite{Grumiaux2021b}.

% Why SRP
Choosing a localization method from all the available methods depends on the type of available acoustic and computational resources, assumptions about the localization scene, and knowledge of the method's mathematical formulation. SRP is known for its straightforward formulation and robust performance in many realistic environments \cite{Zhang2008a}. A historical disadvantage of the method has been its significant computational complexity, although this is of diminishing importance due to the increased computational capacity of today's devices and to the many optimized modifications of \ac{SRP} which have been developed. This has resulted in \ac{SRP} becoming a standard \ac{SSL} method in the literature.

Besides reducing its computational complexity, dozens of \ac{SRP} variants have been developed to improve aspects of its performance, including increasing its robustness in adverse environments or in specific scenarios, and allowing multiple sources or moving sources to be localized. \ac{SRP} can also be used as a feature extractor for neural-based localizers \cite{Diaz-Guerra2021a}. Therefore, one must not only choose \ac{SRP} as a localizer, but must also decide which of the multiple \ac{SRP} `flavours' to use. A prominent flavour is the SRP-PHAT method, which uses the \ac{GCC-PHAT} \cite{Knapp1976b} method as its correlation function, which is shown to offer advantages to other correlation functions for processing speech signals. Unless stated otherwise, the term SRP refers to SRP-PHAT throughout this work.

The goal of this paper is to provide a centralized resource for \ac{SRP} research, to be used by both newcomers and experienced practitioners in the field of \ac{SSL}. Over 200 papers are classified, described and compared, followed by the developement of a modular description of the algorithm, which can be used to develop implementations. A code library named X-SRP is also released as part of this work, with the goal of facilitating the usage of the algorithm. The remainder of the paper comprises the following sections:

\begin{enumerate}
    \setcounter{enumi}{1}
    \item \hyperref[sec:base-model]{\textit{The conventional SRP model}}, which presents \ac{SRP} along with the relevant acoustics concepts required for its comprehension.
    \item \hyperref[sec:computational]{\textit{Reducing SRP's complexity and computational time}}, which discusses papers that focus on reducing \ac{SRP}'s computational cost at a minimal decrease in localization performance.
    \item \hyperref[sec:robustness]{\textit{Increasing robustness}}, which focuses on improving \ac{SRP}'s performance on reverberant and noisy environments using, for example, neural network methods.
    \item \hyperref[sec:multi-source]{\textit{Multi-source SRP approaches}}, which generalizes the conventional \ac{SRP} definition to the detection and localization of multiple simultaneously active sound sources.
    \item \hyperref[sec:applications]{\textit{Practical considerations}}, which include practical applications involving \ac{SRP}, adaptations of the method to track moving sources, to exploit and estimate source and microphone directivity, and comparisons to alternative \ac{SSL} methods.
    \item \hyperref[sec:xsrp]{\textit{X-SRP}}, where a modular description of \ac{SRP} is provided by decomposing the algorithm into functional building blocks. Each of the reviewed papers usually modify a single block in the proposed framework, allowing works to be combined and altered. We apply the created framework by releasing an open-source Python implementation of \ac{SRP} denoted X-SRP, or eXtensible-SRP, with the goal of facilitating collaboration in the field. The released code \footnote{\url{https://github.com/egrinstein/xsrp}} includes implementations of many popular \ac{SRP} variants.   
    \item \hyperref[sec:conclusion]{\textit{Conclusion}}, where a discussion of future research directions is provided and the work is concluded.
\end{enumerate}

\section{The conventional SRP model} \label{sec:base-model}

% History, formulations
The earliest descriptions of \ac{SRP} were provided by Omologo et. al. \cite{Omologo1994a, Omologo1996a, Omologo1997a} and Dibiase et. al. \cite{DiBiase2000a, DiBiase2001b}. Earlier works on \ac{SRP} have also referred to the method as \acf{GCF} \cite{Brutti2005a, Brutti2006a, Brutti2007a, Brutti2008c, Brutti2010a}. The method was later generalized as a \ac{SLF} \cite{Aarabi2003a}. The term \ac{SRP} comes from its guiding principle of searching, or \textit{steering} towards a location which maximizes the output power of a beamformer applied to the microphone signals. Alternatively, \ac{SRP} can also be defined for each pair of microphone signals as the projection of their cross-correlation function in space. Due to its increased clarity, the latter formulation is adopted in this paper. 

This section starts by defining the scope of the problem, followed by the signal model used throughout this paper. Finally, a description of \ac{SRP}'s base model as presented in \cite{DiBiase2000a, DiBiase2001b} is provided. Two alternative formulations are presented, the first in the time domain and the second in the frequency domain, as both are commonly encountered in the literature.

\subsection{Problem statement and definitions}

% Intro
The goal of a localization method is to estimate the positions of one or more sound sources located in space, often an indoor environment. This section focuses on the scenario where a single, static and omnidirectional source located at $\mathbf{u} = [u^{(1)} \, u^{(2)} \, u^{(3)}]^T$ emits a signal $s(t)$ at time $t$; the case of directive, moving and multiple sources are respectively discussed in \autoref{sec:directivity}, \autoref{sec:tracking} and \autoref{sec:multi-source}. The source can also be expressed in spherical coordinates $\mathbf{u} = [\phi \, \theta \, \rho]^T$ with respect to a reference point, typically the centre of a microphone array. Variables $\phi$, $\theta$ and $\rho$ respectively represent the source's \textit{azimuth}, \textit{elevation} and \textit{range}. The source locations are estimated using signals $x_m(t)$ received from an array of $M$ microphones, each located at known positions $\mathbf{v}_m = [v_m^{(1)} \, v_m^{(2)} \, v_m^{(3)}]^T$, $m=1,\,\hdots,\,M$.

\subsubsection{Near- versus Far-field localization} \label{sec:near-far-field}

% PSSL, distributed arrays and 
This subsection discusses the different types of localization which are frequently encountered in the literature, namely, \acf{PSSL} and \ac{DOA} estimation. \ac{PSSL} consists of fully estimating the source's position, and is usually employed when the distances between microphones in the array is similar to the distance between the microphones and the source. This is equivalent to saying the source is located in the \textit{near-field} of the array. This configuration is referred to as a \textit{distributed} array, which can be constituted for example of multiple network-connected devices such as laptops, cell phones or voice assistants. In this case, as each device has their own \ac{ADC}, they must be synchronized to a common sampling frequency $f_s$, or a compensation algorithm must be applied to the signals to prevent synchronization issues \cite{Chinaev2019}.

Conversely, when employing a centralized microphone array such as a single voice assistant, the distance between microphones is usually significantly smaller than the distance between the sources of interest and the array itself. This is equivalent to saying the source is located in the array's \textit{far-field}. In this case, the spherical wave leaving the source is observed as a plane wave which has no defined origin: an infinite set of sources may produce a plane wave with the same incident angle to the array. For this reason, the range $\rho$ is usually not estimated when using compact arrays. The task of estimating the azimuth, $\phi$, and elevation, $\theta$, is referred to as \ac{DOA} estimation. 

\subsection{Signal model}

The received signal $x_m(t)$ at microphone $m$ is equal to 
\begin{equation} \label{eq:prop-reverb}
    x_m(t) = \int_{-\infty}^{\infty} h_m(r; \mathbf{u}) s(t - r) \mathrm{d}r + \epsilon_m(t),
\end{equation}
that is, a convolution between the source signal $s(t)$ and a \acf{RIR} $h_m(r; \mathbf{u})$, which models the propagation effects and reverberation, plus a noise term $\epsilon_m(t)$. However, \ac{SRP} adopts a simplified propagation model where reverberation is modeled using the noise term $\epsilon_m(t)$. This free-field is defined as
\begin{equation} \label{eq:prop-anechoic}
    x_m(t) = a_m(\mathbf{u}) s(t - \tau_m(\mathbf{u})) + \epsilon_m(t) ,
\end{equation}
that is, the signal emitted by the source is received at microphone $m$ attenuated by a factor $a_m(\mathbf{u})$, delayed by $\tau_m(\mathbf{u})$ seconds and corrupted by a measurement noise term $\epsilon_m(t)$. This is equivalent to adopting the \ac{RIR} in \eqref{eq:prop-reverb} as a pure impulse $h_m(t; \mathbf{u}) = a_m(\mathbf{u}) \delta(t - \tau_m(\mathbf{u}))$. Note that this model assumes attenuation to be frequency-independent. The attenuation and delay effects will be further detailed in \autoref{sec:acoustics}.

Alternatively, it is often advantageous to define \eqref{eq:prop-anechoic} in the time-frequency domain, by decomposing the source signal into complex-valued sinusoids $s(t, f)$ of frequencies $f$. In practice, such a signal can be obtained by applying the Fourier transform on $s(t)$. The received signal $x_m(t, f)$ is then defined for each time-frequency pair $(t, f)$ as
\small
\begin{equation} \label{eq:prop-freq}
    x_m(t, f) = s(t, f)a_m(\mathbf{u}, f)e^{-jf\tau_m(\mathbf{u})} + \epsilon_m(t, f).
\end{equation}
\normalsize
The advantage of \eqref{eq:prop-freq} in comparison to \eqref{eq:prop-anechoic} is that delay, $\tau_m$, and attenuation, $a_m(f)$, effects can be jointly represented by multiplication with a signal complex-valued scalar.

Although the above definitions are conceptually useful, in practice, \ac{SRP} is computed using a \textit{frame} or vector of dimension $L$ samples for each microphone. A frame $\mathbf{x}_m(t)$ is defined in the time domain as
\small
\begin{equation}
    \mathbf{x}_m(t) = [x_m(t) \, x_m(t-T_s) \, ... \, x_m(t-(L-1)T_s)]^T, \label{eq:frame_t}
\end{equation}
\normalsize
where $T_s = 1/f_s$. Furthermore, a frequency domain frame $\bar{\mathbf{x}}_m(t)$ is defined as
\begin{equation}
    \bar{\mathbf{x}}_m(t) = \text{DFT}(\mathbf{x}_m(t)), \label{eq:frame_f}
\end{equation}
that is, the application of the \ac{DFT} to temporal frame $\mathbf{x}_m(t)$. $\bar{\mathbf{x}}_m(t)$, where each of its entries represents a time-frequency bin $\bar{x}_m(t,f)$ with $f \in \mathcal{F}$, where
\small
\begin{equation}
    \mathcal{F} = \{f| f=-f_s/2 + kf_s/L, \, k=0,...,L-1\}, \label{eq:analysis-freqs}
\end{equation}
\normalsize
constitutes the set of analysis frequency components used.

\subsection{Acoustics, TOF and TDOA} \label{sec:acoustics}
In this subsection, we further contextualize the signal model defined in \eqref{eq:prop-anechoic} and \eqref{eq:prop-freq} using relevant acoustic principles.

A sound wave emanating from the source location $\mathbf{u}$ travels at the speed of sound $c$ to each microphone's location $\mathbf{v}_m$. The propagation time $\tau_m(\mathbf{u})$, also known as the \acf{TOF} between the source at $\mathbf{u}$ and microphone $m$, can therefore be expressed, in seconds, as
\begin{equation} \label{eq:propagation_time_distance}
    \tau_m(\mathbf{u}) = \frac{\lVert \mathbf{u} - \mathbf{v}_m \rVert}{c}.
\end{equation}
If $\tau_m(\mathbf{u})$ can be correctly estimated for three or more microphones, an estimate of $\mathbf{u}$ can be obtained. This is the strategy used by \textit{active} localization systems \cite{So2011c}, which use controlled and/or known source signals so that the emission time of the source signal is accessible. Conversely, \ac{SRP} is a \textit{passive} localization method which allows for a broader range of sources, such as human speakers, to be localized. In the case of compact arrays where the near-field assumption holds \eqref{sec:near-far-field}, the TDOA definition in \eqref{eq:tdoa} can be approximated as
\begin{equation} \label{eq:tdoa-ff}
    \tilde{\tau}_{lm}(\mathbf{u}) = \frac{(\mathbf{v}_l - \mathbf{v}_m)^T}{c} . \frac{\mathbf{u}}{\lVert \mathbf{u} \rVert},
\end{equation}
that is, the dot product between the vector $\mathbf{v}_l - \mathbf{v}_m$ and the normalized source direction $\frac{\mathbf{u}}{\lVert \mathbf{u} \rVert}$, scaled by $1/c$.

\begin{figure}
\centering
    \def\tikzscale{0.18}

\begin{tikzpicture}[scale=\tikzscale]

  \tikzset{
      elli/.style args={#1:#2and#3}{
          draw,
          shape=ellipse,
          rotate=#1,
          minimum width=2*#2,
          minimum height=2*#3,
          outer sep=0pt,
      }
  }
  
  %
  % #1 optional parameters for \draw
  % #2 angle of rotation in degrees
  % #3 offset of center as (pointx, pointy) or (name-o-coordinate)
  % #4 length of plus (semi)axis, that is axis which hyperbola crosses
  % #5 length of minus (semi)axis
  % #6 how much of hyperbola to draw in degrees, with 90 you’d reach infinity
  %
  \newcommand\tikzhyperbola[6][thick]{%
      \draw [#1, rotate around={#2: (0, 0)}, shift=#3]
          plot [variable = \t, samples=1000, domain=-#6:#6] ({#4 / cos( \t )}, {#5 * tan( \t )});
      \draw [#1, rotate around={#2: (0, 0)}, shift=#3]
          plot [variable = \t, samples=1000, domain=-#6:#6] ({-#4 / cos( \t )}, {#5 * tan( \t )});
  }

% Command for plotting one of the curves (left)
    \newcommand\tikzhyperbolanew[6][thick]{%
      \draw [#1, rotate around={#2: (0, 0)}, shift=#3]
          plot [variable = \t, samples=1000, domain=-#6:#6] ({-#4 / cos( \t )}, {#5 * tan( \t )});
  }
  
  \def\angle{0}
  \def\bigaxis{3.77cm}
  \def\smallaxis{3.2cm}
  
  \coordinate (center) at (0,0);
  \coordinate (v1) at (-5, 0);
  \coordinate (v2) at (5, 0);
  \coordinate (u) at (-10.15, -8);
  \coordinate (circle1) at (-9.45, 0);
  % \node [scale=\tikzscale, elli=\angle:\bigaxis and \smallaxis, line width = 1.2pt, color=black, dotted] at (center) (e) {};
  \node [scale=\tikzscale] at (center) (e) {};
  \draw [-{stealth}, line width = 1.2pt, color = black] ([shift={(\angle:-20)}] e.center) -- ([shift={(\angle:20)}] e.center) node [above right] {$ $}; % {u^x};
  \draw [-{stealth}, line width = 1.2pt, color = black] ([shift={(90+\angle:-20)}] e.center) -- ([shift={(90+\angle:20)}] e.center) node [above left]  {$ $}; % {u^y};
  \tikzhyperbolanew[line width = 1.2pt, color=blue!80!black]{\angle}{(center)}{\bigaxis}{\smallaxis}{77}
 % \draw [line width = 1.2pt, color = red, dashed] (-19.75,-16.47) -- (-1.05,0) node [above right] {}; % dashed line top
 % \draw [line width = 1.pt, color = red, dashed] (-19.75,16.47) -- (-1.05,0) node [above right] {};% dashed line bottom
  
  %\draw [gray] (circle1) circle (8.019cm);

  \node[circle,fill=orange,inner sep=0pt,minimum size=5pt,label=135:{$\mathbf{v}_1$}] (mic1) at (v1) {};

  \node[circle,fill=orange,inner sep=0pt,minimum size=5pt,label=45:{$\mathbf{v}_2$}] (mic2) at (v2) {};

  \node[circle,fill=blue,inner sep=0pt,minimum size=5pt,label=below:{$\mathbf{u}$}] (s1) at (u) {};

  % \pgfmathsetmacro\axisratio{\smallaxis / \bigaxis}
  
  % asymptotes
  % \def\lengthofasymptote{15}
  % \draw [color=black!40, line width = 0.4pt, rotate around={\angle + atan( \axisratio ): (center)}]
  %     ($ (-\lengthofasymptote, 0) + (center) $) -- ++(2*\lengthofasymptote, 0) ;
  % \draw [color=black!40, line width = 0.4pt, rotate around={\angle - atan( \axisratio ): (center)}]
  %     ($ (-\lengthofasymptote, 0) + (center) $) -- ++(2*\lengthofasymptote, 0) ;
  
  % \tikzhyperbola[line width = 1.2pt, color=red!80!black]{90+\angle}{(center)}{\smallaxis}{\bigaxis}{76}
  
  \end{tikzpicture}
  \caption{Hyperbola branch of points with the same TDOA as a source located at $\mathbf{u}$ with respect to microphone positions $\mathbf{v}_1$ and $\mathbf{v}_2$.}
  \label{fig:hyper}
\end{figure}
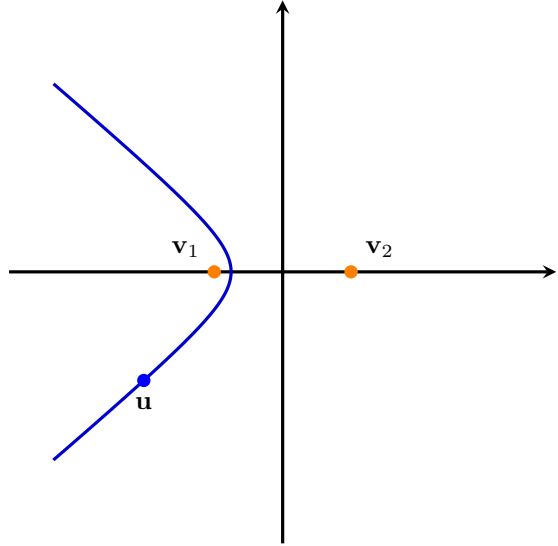

\ac{SRP} performs passive localization by exploiting the \textit{relative} delay, also known as the \acf{TDOA}, between \textit{pairs} of microphones. The importance of the \ac{TDOA} and its relationship with the cross-correlation function between pairs of signals will be discussed in detail in \autoref{sec:cc-tdoa}. Using \eqref{eq:propagation_time_distance}, $\tau_{lm}$ is defined, in seconds, as
\small
\begin{equation} \label{eq:tdoa}
    \tau_{lm}(\mathbf{u}) = \tau_l(\mathbf{u}) - \tau_m(\mathbf{u}) = \frac{\lVert \mathbf{u} - \mathbf{v}_l \rVert - \lVert \mathbf{u} - \mathbf{v}_m \rVert}{c}.
\end{equation}
\normalsize
The \ac{TDOA} for a pair of microphones can be interpreted as how much earlier/later a signal arrives at the first microphone in comparison to the \ac{TOA} at the second microphone. Multiple positions $\mathbf{u}$ can produce the same delay $\tau_{lm}$ for a pair of microphones fixed at $(\mathbf{v}_l, \mathbf{v}_m)$. These positions lie along a hyperbola/hyperboloid branch in 2D/3D, as shown in \cite{Gustafsson2003a} and can be viewed in \autoref{fig:hyper}. The maximum possible \acp{TDOA} for a microphone pair occurs when the source and microphones are collinear and the source is not located between the microphones, and has an absolute value of
\begin{equation} \label{eq:max_tdoa}
        \left| \tau_{lm}^{\text{lim}} \right| = \frac{\lVert \mathbf{v}_l - \mathbf{v}_m \rVert}{c}.
\end{equation}
By determining the intersection of the hyperbolas produced by multiple microphone pairs, the source position, can be estimated as $\hat{\mathbf{u}}$. Approaches utilizing this strategy are known as \textit{triangulation}, \textit{TDOA-based}, \textit{indirect} or \textit{two-step} approaches \cite{So2011c}, since they require a first step of estimating the \acp{TDOA} before a second step of estimating the source locations. Although these approaches are less computationally expensive than \ac{SRP}, their reliance on the estimated \acp{TDOA} make them non-robust in adverse noisy or reverberant scenarios \cite{Dmochowski2007e}.

\subsection{Estimating TDOA: Cross-correlation and GCC-PHAT} \label{sec:cc-tdoa}

\begin{figure}
\centering
    \includegraphics[width=\columnwidth]{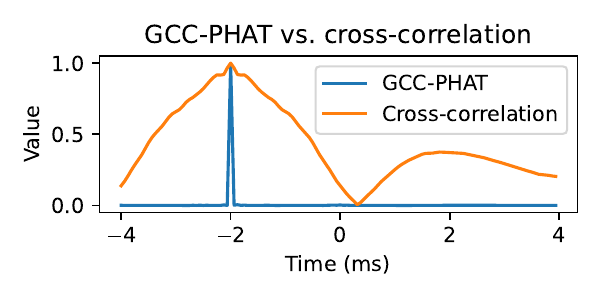}
  \caption{Example comparison between the normalized temporal cross-correlation and GCC-PHAT for a scenario containing two microphones and a source producing a speech signal with a TDOA of -2~ms.}
  \label{fig:gcc-vs-cc}
\end{figure}

% Cross-correlation
The \ac{TDOA} $\tau_{lm}$ between two microphones can be estimated as the argument of the peak of the cross-correlation between microphone signal frames $\mathbf{x}_l(t)$ and $\mathbf{x}_m(t)$. The discrete cross-correlation (CC) function is defined as 
\begin{align}
    \text{CC}(\tau ; \mathbf{x}_l, \mathbf{x}_m) &= \sum_{n=0}^{L-1} 
    x_l(t)[n]x_m(t-nT_s-\tau) \nonumber \\
    &= \mathbf{x}_l^T(t)\mathbf{x}_m(t-\tau), \label{eq:cc-time}
\end{align}
where $\tau$ must be a multiple of the sampling period $T_s$ and appropriate zero padding is applied. Despite its straightforward formulation, \eqref{eq:cc-time} is seldom used in practice for localizing speech sources in reverberant and noisy environments, as the non-flat spectrum of the source signal reduces the selectivity of the function. Instead, the \acf{GCC-PHAT} function \cite{Knapp1976b, Omologo1997a} is usually adopted. `Generalized' comes from the fact that a cross-correlation value is produced for every frequency component of the signals after a pre-filtering operation. This operation is typically the `Phase Transform' weighting, which whitens the frequency components, thus sharpening the correlation peak. The \ac{GCC-PHAT} function is defined for frequency-domain microphone signals  $\bar{\mathbf{x}}_l(t)$ and $\bar{\mathbf{x}}_m(t)$, typically processed frame-by-frame, as
\small
\begin{equation} \label{eq:gcc-phat}
    \text{GCC-PHAT}(f ; \bar{\mathbf{x}}_l, \bar{\mathbf{x}}_m) = \frac{\bar{\mathbf{x}}_l(t,f)\bar{\mathbf{x}}^{*}_m(t,f)}{\left| \bar{\mathbf{x}}_l(t,f)\right| \left|\bar{\mathbf{x}}_m(t,f) \right|}.
\end{equation}
\normalsize
The phase transform is applied using the denominator of \eqref{eq:gcc-phat}. In practice, \eqref{eq:gcc-phat} is computed for a set of analysis frequencies $\mathcal{F}$ to generate a GCC \textit{frame} $\bar{\mathbf{g}}$. By applying the \ac{IDFT}, a time-domain time-domain vector $\mathbf{g}$ can obtained,
\begin{equation} \label{eq:gcc-time-frame}
    \mathbf{g} = \text{IDFT}(\bar{\mathbf{g}}),
\end{equation}
where each entry $\mathbf{g}[k]$ represents a temporal correlation value between $\mathbf{x}_l$ and $\mathbf{x}_m$ at sample $k$. A frame can be built in a similar manner using the temporal CC in \eqref{eq:cc-time}. An example comparison between two frames computed using \eqref{eq:gcc-time-frame} and temporal $\text{CC}$ is shown in \autoref{fig:gcc-vs-cc}, where it can be observed that the peak produced by GCC-PHAT is much sharper than by $\text{CC}$. 

In an ideal scenario, the temporal CC or \\ \ac{GCC-PHAT} function exhibits a sharp peak at $\tau_{lm}$, which can be used for two-step methods. However, under reverberant or noisy scenarios, the cross-correlation function can exhibit multiple peaks, rendering the TDOA estimates and the subsequent triangulation-based approaches unreliable. As we will show in the following section, \ac{SRP} applies the principle of least commitment \cite{Dmochowski2007e, Birchfield2001a}; instead of estimating $\tau_{lm}$ early on and discarding all other values and peaks of the cross-correlation function, \ac{SRP} associates each cross-correlation value with a candidate locus in space using \eqref{eq:tdoa}. 

\subsection{Time-domain SRP formulation}
The conventional \ac{SRP} for a candidate source location $\mathbf{u}$ and a pair of microphones $(l, m)$ is defined as 
\cite{DiBiase2000a, DiBiase2001b}
\begin{equation} \label{eq:pairwise_srp}
    \text{SRP}_{lm}(\mathbf{u} \, ; \,\, \mathbf{x}_l, \mathbf{x}_m) = \text{CC}(\lfloor \tau_{lm}( \mathbf{u}) \rceil ; \mathbf{x}_l, \mathbf{x}_m),
\end{equation}
that is, the cross-correlation function between signal frames $\mathbf{x}$ and $\mathbf{x}_m$, evaluated at delay $\lfloor \tau_{lm}( \mathbf{u}) \rceil$, where $\lfloor \cdot \rceil$ represents rounding to the nearest multiple of $T_s$. Note that the time index $t$ is hereafter omitted for clarity, and that time-domain $\text{GCC-PHAT}$ defined in \eqref{eq:gcc-time-frame} is usually preferred to \eqref{eq:cc-time} in practice for its improved performance on realistic scenarios. We the time-domain SRP formulation only for the sake of completeness. Finally, the global \ac{SRP} is defined as the sum of all pairwise SRPs,
\begin{equation} \label{eq:srp}
    \text{SRP}(\mathbf{u} \, ; \,\, \mathcal{X}) = \sum_{l=1}^M \sum_{m=l+1}^{M} \text{SRP}_{lm}(\mathbf{u} \, ; \,\, \mathbf{x}_l, \mathbf{x}_m),
\end{equation}
where $\mathcal{X} = \{\mathbf{x}_1,\, ...,\, \mathbf{x}_M \}$ is the set of $M$ \mbox{$L$-dimensional} frames pertaining all microphones. This value is related to the likelihood of a source being located at a candidate point $\mathbf{u}$. The complete \ac{SRP} method consists of evaluating \eqref{eq:srp} for a set of candidate locations and selecting the location maximizing \eqref{eq:srp} as the estimated location. The set of candidate locations typically consists of a regularly sampled spatial grid. The grid construction procedure will be defined in \autoref{sec:grid}.

\subsection{Frequency-domain SRP formulation}
This formulation decomposes the microphone signals into frequency bands, which are independently analysed using \ac{GCC-PHAT} in \eqref{eq:gcc-phat} as
\begin{equation} \label{eq:freq-pairwise-srp-phat}
    \begin{split}
    \text{SRP}_{lm}(\mathbf{u}, & f \, ; \,\, \bar{\mathbf{x}}_l, \bar{\mathbf{x}}_m) = \\
                                       & \text{GCC-PHAT}(f ; \bar{\mathbf{x}}_l, \bar{\mathbf{x}}_m) e^{jf\tau_{lm}(\mathbf{u})}.
    \end{split}
\end{equation}
Equation \eqref{eq:freq-pairwise-srp-phat} can be interpreted as steering, or shifting, the microphone signal $x_m(f)$ by a phase $jf\tau_{lm}(\mathbf{u})$. Note that although \eqref{eq:freq-pairwise-srp-phat} produces a complex-value, its imaginary part is typically discarded as irrelevant \cite{Dietzen2021a}. Finally, the global \ac{SRP} is represented in the frequency domain in a similar way to the time-domain formulation \eqref{eq:srp}, after summing over the set $\mathcal{F}$ of frequencies being analyzed,
\small
\begin{equation}
    \label{eq:srp_freq}
    \text{SRP}(\mathbf{u} \, ; \,\, \bar{\mathcal{X}}) = \sum_{l=1}^M \sum_{m=l+1}^{M} \sum_{f \in \mathcal{F}} \text{SRP}_{lm}(\mathbf{u}, f \, ; \,\, \bar{\mathbf{x}}_l, \bar{\mathbf{x}}_m),
\end{equation}
\normalsize
where $\bar{\mathcal{X}}$ is the frequency-domain representation of $\mathcal{X}$. In practice, the frequencies used may be lower than the Nyquist rate to prevent a phenomenon called spatial aliasing \cite{Dmochowski2009}. Note the time and frequency definitions of SRP are not equivalent due to this aforementioned low-pass filtering, as well as the rounding operator required when constructing a temporal CC or GCC vector makes \eqref{eq:srp} operate using integer delays, which may not correspond to the true source's TDOA. The error due to rounding may be reduced by using distributed arrays or mitigated by applying interpolation \cite{Tervo2008a}. However, significant errors may be produced for compact arrays, where the \ac{TDOA} range defined by \eqref{eq:max_tdoa} is typically only a few samples \cite{Nunes2012}.  

\subsection{Grid construction and search} \label{sec:grid}

To estimate the location of the source, \eqref{eq:srp} or \eqref{eq:srp_freq} are evaluated over a set, $\mathcal{G}$, of $G$ candidate positions relative to a reference point in the room, typically one of its corners. The elements of $\mathcal{G}$ are usually defined by creating a uniform spatial grid. For performing \ac{SSL} in a cuboid-shaped room, a cuboid-shaped grid is typically used. For example, when performing planar or 2D localization $\lvert \mathcal{G} \rvert = \lvert \mathcal{G}^{(1)} \rvert \times \lvert \mathcal{G}^{(2)} \rvert$, where $\lvert \mathcal{G}^{(1)} \rvert$ and $\lvert \mathcal{G}^{(2)} \rvert$ are respectively the number of points used for the width and length dimension. $\mathcal{G}$ becomes
\begin{equation} \label{eq:cartesian-grid}
\begin{split}
    \mathcal{G}_{2D} =
    \; \{ \; [g^{(1)} &R^{(1)} \,\, g^{(2)} R^{(2)}]^T  \mid \\ 
        & g^{(1)} \in \{1,\ldots,\lvert \mathcal{G}^{(1)}_{2D} \rvert\} \\
        & g^{(2)} \in \{1,\ldots,\lvert \mathcal{G}^{(2)}_{2D} \rvert\} \},
\end{split}
\end{equation}
where $R^{(1)} = D^{(1)}/\lvert \mathcal{G}^{(1)}_{2D} \rvert$ and $R^{(2)} = D^{(2)}/\lvert \mathcal{G}^{(2)}_{2D} \rvert$ are the width and length \textit{resolution} for a room of width $D^{(1)}$ and length $D^{(2)}$. Conversely, when performing planar or 2D \ac{DOA} estimation, the grid can be made by setting the origin to the microphone array centre, and a circular grid is created,

\small
\begin{equation} \label{eq:doa-grid}
    \begin{split}
        \mathcal{G}_\text{DOA2D} & =  \\
        & \{ \; [\cos(\phi) \,\, \sin(\phi])]^T \mid \phi \in \\
        & \{R^{(\phi)}, 2R^{(\phi)} \ldots, 2\pi\} \}.
    \end{split}
\end{equation}
\normalsize
In \eqref{eq:doa-grid}, each point represents a distinct candidate source direction. Furthermore, neighboring points are separated by the angular resolution $R^{(\phi)}$, where $\phi$ is the candidate source's \textit{azimuth}. In 3D \ac{DOA} estimation, the \textit{elevation}, defined as the angle between the segment connecting the source and array centre and the horizontal plane is also estimated.

For both \ac{DOA} estimation and \ac{PSSL}, the complete \ac{SRP} map consists of evaluating the SRP function for all candidate locations in the grid $\mathcal{G}$, and selecting the location producing the maximum SRP value as the estimated position,
\begin{equation} \label{eq:vanilla-grid-search}
    \hat{\mathbf{u}} = \argmax_{\mathbf{u} \in \mathcal{G}} \text{SRP}(\mathbf{u}).
\end{equation}

\begin{figure}
\centering
   \includegraphics[width=\columnwidth]{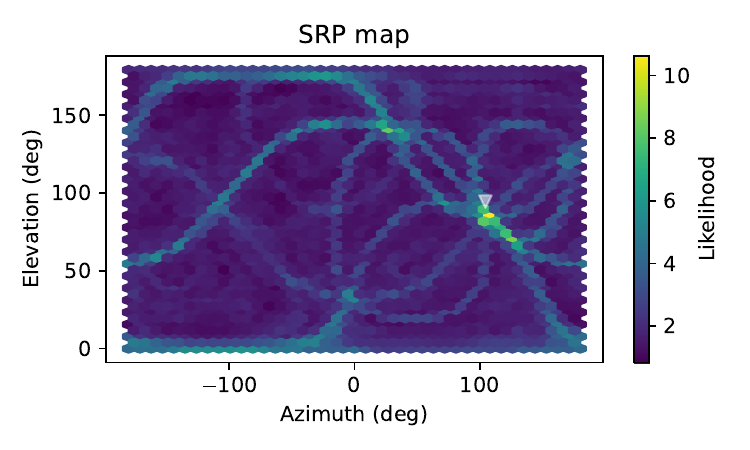}
  \caption{Example of an SRP map for the task of 3D DOA estimation of a speech source using a spherical array of 8 microphones. Reverberation was simulated with a reverberation time of $T_{60} = 400$~ms, and the source is located below the transparent triangle at $(100^o, 60^o)$. Spatially uncorrelated white noise was added to the microphones at 20~dB SNR.}
  \label{fig:srp-doa}
\end{figure}

An example of an \ac{SRP} map for a simulated environment of low reverberation is shown in \autoref{fig:srp-doa}.

\section{Reducing SRP's complexity and computational time} \label{sec:computational}
\subsection{Complexity analysis}
We start by outlining the computational complexity of the frequency-domain, conventional SRP method as defined in \eqref{eq:srp_freq}. Here, complexity is measured by the number of real multiplications and divisions performed by the algorithm, ignoring the additions, as commonly done. Furthermore, we follow the  Bachmann–Landau (or big-$O$) notation, which measures asymptotic behaviour of algorithm complexity as input sizes grow.

The method can be divided into four sequential operations. The first two operations consist of extracting the \ac{DFT} for each frame of the $M$ microphones followed by computing the \ac{GCC-PHAT} for all $P$ microphone pairs, where $P = M(M-1)/2$. In practice, the \ac{FFT} algorithm \cite{Cochran1967} is used to implement the \ac{DFT}. The \ac{FFT} has a complexity of $\text{O}(L \log L)$ . We assume the \ac{FFT} operation converts a time-domain frame of size $L$ into a frequency-domain frame of same size. Since \ac{GCC-PHAT} consists of an element-wise multiplication of the vectors $\bar{\mathbf{x}}_l$ and $\bar{\mathbf{x}}_m$ divided by their respective magnitudes, its complexity is therefore  $\text{O}(L)$. 

The third step is the creation of the $P$ pairwise \ac{SRP} likelihood grids of size $G=\lvert \mathcal{G} \rvert$, for all $L$ frequencies, followed by their sum to create a global \ac{SRP} grid. As this operation consists of multiplying the \acp{GCC-PHAT} by an exponential $e^{jf\tau_{lm}(\mathbf{u})}$, its complexity is $\text{O}(GPL)$. The final step consists of comparing all grid points to obtain the argument of its maximum, which is the estimated source location. As comparisons are often assumed to offer a lower complexity, this last step is ignored. The number of operations performed by \ac{SRP} is thus obtained as
\begin{equation} \label{eq:csrp-complexity}
    \begin{split}
        \text{O}_{\overline{\text{SRP}}} & = \text{O}\big( M L \log L + PL + GPL \big), \\
        & \simeq \text{O}\big( M L \log L + GPL \big),
    \end{split}
\end{equation}
where the three terms in the first line represent each of the sequential operations discussed above. The simplification on the bottom line is obtained by removing the second term, as $G \gg 1$. We can see from \eqref{eq:csrp-complexity} that straightforward strategies can be followed to reduce the complexity of \ac{SRP}. One is to use only a subset of microphones $M' < M$ or subselecting $P' < M(M - 1)/2$ pairs instead of evaluating all pair combinations. Another is to employ a smaller frame size $L$ and reducing the frequency range in which the SRP map is computed. Finally, a coarser grid can be employed. All these strategies come, however, with a reduction in localization performance. Most of the research presented in this section proposes strategies to reduce the grid size $G$, or modify the functionality of the conventional \ac{SRP} method while minimizing the loss in localization performance. 

In turn, the computational complexity of time-domain SRP in \eqref{eq:srp} is smaller than in \eqref{eq:csrp-complexity}, as a single map is computed in the time domain instead of $L$ frequency domain maps, i.e., it uses one less nested `for each' loop. The complexity of \eqref{eq:srp} is therefore expressed as
\begin{equation} \label{eq:csrp-complexity-time}
    \begin{split}
        \text{O}_\text{SRP} & = \text{O}\big( M L \log L + PL + GP \big),
    \end{split}
\end{equation}
which disregards the negligible inverse \ac{DFT} used to obtain the temporal GCC vector \eqref{eq:gcc-time-frame}. Furthermore, projection of the cross-correlation function is achieved in \eqref{eq:srp} by accessing an element in the cross-correlation vector, which is more computationally efficient, albeit less precise, than the multiplication by a complex exponential used in the frequency-domain version.

\subsection{Coarse grids and Volumetric-SRP}

\begin{figure} 
\centering
    \includegraphics[width=0.9\columnwidth]{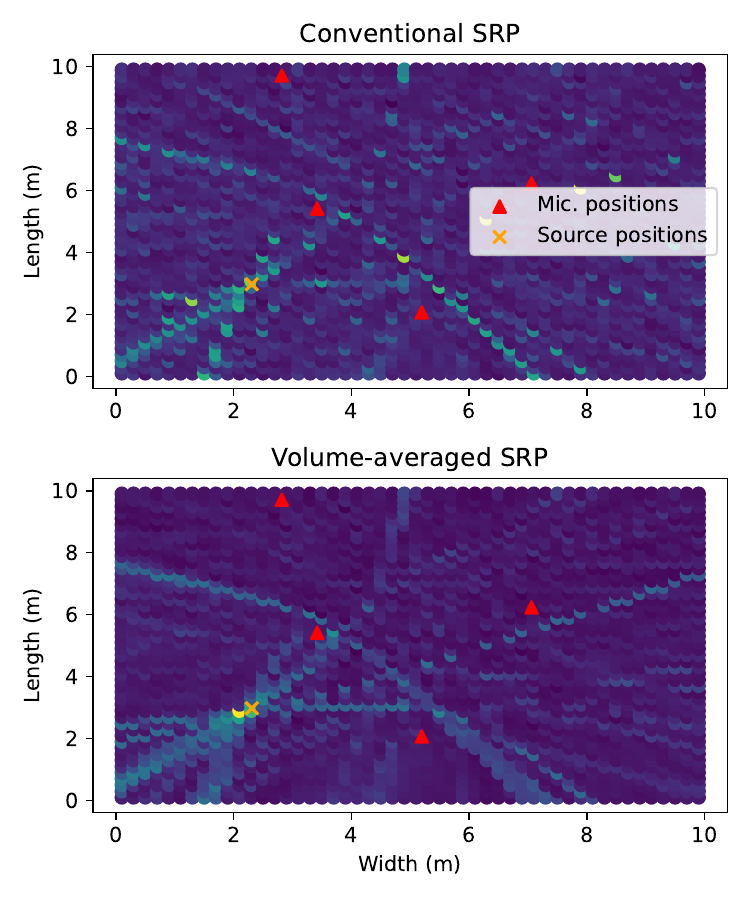}
  \caption{Comparison between SRP maps generated with (bottom) and without (top) volumetric techniques.}
  \label{fig:srp-volumetric}
\end{figure}

As mentioned above, reducing $G$ is a straightforward strategy for reducing \ac{SRP}'s complexity. When applying equispaced grids such as those described in \eqref{eq:cartesian-grid} and \eqref{eq:doa-grid}, this can be achieved by reducing the resolution parameters $R^{(1)}$, $R^{(2)}$ and $R^{(\phi)}$. However, this comes with the risk of not sampling the true source location, which may lead to the peak of the cross-correlation function not to be projected into the map, leading to a high localization error \cite{Garcia-Barrios2021a}. Nonetheless, many strategies can be applied to increase the localization performance of approaches using coarse grids.

% Volumetric 
As grids become coarser, each point is associated with an increasingly larger spatial region or volume. It is therefore reasonable to devise a way to modify \ac{SRP}'s operation to take into account the entire set of points around the candidate. Methods employing this strategy are referred to as \textit{volumetric} SRP (V-SRP) \cite{Nunes2014a, Lima2015b, Lima2015c}. An example comparison between conventional and volumetric \ac{SRP} maps is shown in \autoref{fig:srp-volumetric}. The volume surrounding a candidate position is defined as
\begin{equation}
    \begin{split}
        \mathcal{V}(\mathbf{u}) = \{ \\
        &[x \; y \; z]^T \; \vert \; \\
        & | x -  u^{(1)} | \leq r^{(1)}/2 \\
        & | y -  u^{(2)} | \leq r^{(2)}/2 \\
        & | z -  u^{(3)} | \leq r^{(3)}/2 \\
    \},
    \end{split}
\end{equation}
where $r^{(1)}$, $r^{(2)}$ and $r^{(3)}$ respectively represent the width, length and height of the volume. The Volumetric SRP (V-SRP) approach is typically defined by considering the SRP value of all points within the volume, which are then combined using a pooling function such as summation. The pairwise V-SRP function can thus be defined as
\small
\begin{equation} \label{eq:v-srp}
    \text{V-SRP}_{lm}(\mathcal{V}; \mathcal{X}) = \sum_{\tau = \min(\tau_{lm}(\mathbf{u} \in \mathcal{V} ))}^{\max(\tau_{lm}(\mathbf{u}))} \mathbf{g}[\tau ; \mathbf{x}_l, \mathbf{x}_m].
\end{equation}
\normalsize
Different approaches and approximations can be used to find the summation interval in \eqref{eq:v-srp}. A popular approach is the Modified SRP (M-SRP) \cite{Cobos2011a}, which approximates the minimum and maximum \ac{TDOA} limits in the volume by first remarking that, due to the hyperboloidal nature of \acp{TDOA}, the extremes must be contained in the volume's boundary. These values are then approximated using the TDOA's gradient vector and the centre of the volume. Although summation is used as a pooling function in \eqref{eq:v-srp}, it has been shown that average \cite{Marti2013a} or max \cite{Salvati2022a} pooling may increase robustness to noise and reverberation. 

The work of \cite{Nunes2014a} proposes exacts bounds for the maximum and minimum and maximum TDOA limits used in the M-SRP algorithm \cite{Cobos2011a} in anechoic conditions. In particular, the authors show that the minimum and maximum TDOAs of a cuboid volume can be always found by searching a set of only 26 points involving its vertices, edges and faces. Furthermore, this can be further appriximated by searching only the volume’s 8 vertices, further simplifying finding the maximum and minimum TDOAS as these limits can be precomputed for any given cuboid and microphone array locations. The computational complexity of M-SRP \eqref{eq:v-srp} can be further reduced through an iterative subdivision of the maximal volume \cite{Marti2013a, Nunes2014a, Boora2021a, Boora2022c}.

\subsection{Iterative grid refinement} \label{sec:iterative}

A common strategy used in conjunction with coarse grids consists of iteratively modifying the initial search grid $\mathcal{G}(0)$ based on the candidate position's \ac{SRP} values, allowing for the algorithm to `focus' on promising regions. This procedure can be applied repeatedly until a stopping condition is reached, i.e.,
\begin{equation}
    \begin{split}
        \mathcal{G}(i) = \text{ITERATE}(\mathcal{G}(i - 1)),
    \end{split}
\end{equation}
where the $\text{ITERATE}$ function usually involves evaluating the \ac{SRP} function on the current grid candidate points, discarding points based on a criterion, and generating additional candidates based on some heuristic.

This iterative procedure may be performed using a quadtree \cite{Duraiswami2001, Zotkin2004b}, a tree-based data structure commonly used for image processing. In \cite{Zotkin2004b}, each cell of an initial azimuth-elevation square grid of size $16\times16$ is iteratively subdivided into four non-overlapping cells, where the \ac{SRP} function is computed on each region's centre. To prevent the grid size from growing exponentially, only the cells with the highest 
\ac{SRP} value are selected for further division.

\begin{figure} 
\centering
    \includegraphics[width=\columnwidth]{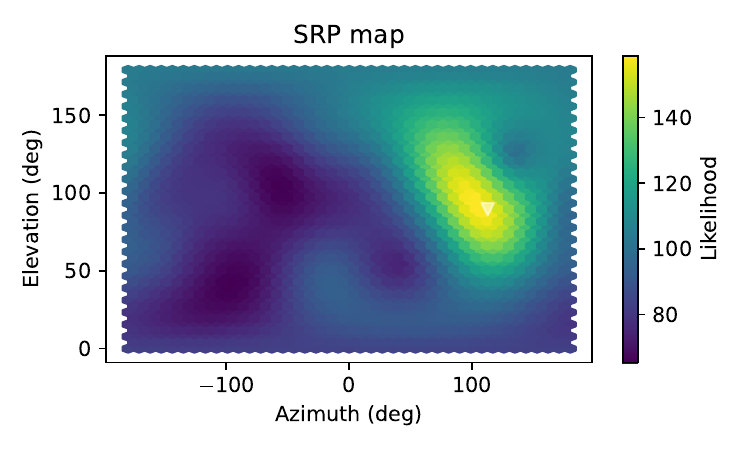}
  \caption{Low-pass version of the frequency-domain SRP, where only frequencies up to 200 Hz are considered.}
  \label{fig:srp_low_pass}
\end{figure}

When a coarse grid is used, the true source location $\mathbf{u}$ may lie on a grid point. A strategy to ensure $\mathbf{u}$'s neighbours exhibit a high \ac{SRP} value in initial iterations was proposed in \cite{Zotkin2004b},
which identify that the width of a peak on an \ac{SRP} map is inversely proportional to the source’s carrier frequency. Therefore, computing \ac{SRP} using only low frequencies produces a smoother map.
This is illustrated in \autoref{fig:srp_low_pass}, where only frequencies below 200~Hz are used for $\mathcal{F}$, which can be compared to \autoref{fig:srp-doa} which shows a map generated using all frequencies up to the Nyquist rate. 

\begin{figure}
\centering
    \includegraphics[width=0.65\columnwidth]{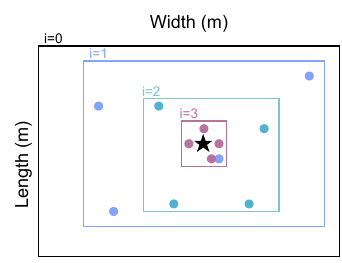}
  \caption{Iterative Region contraction procedure, where different colours represent search regions and grids of points related to iterations i. The true source location is represented by the black star.}
   \label{fig:region_contraction}
\end{figure}

The initial grid can also consist of points randomly sampled on the room's boundaries, as formulated in the \acf{SRC} method defined in \cite{Do2007c}. The region contraction procedure is exemplified in \autoref{fig:region_contraction}. The subsequent grid can be chosen by resampling a set of points on the smaller boundary containing the previous candidates exhibiting the highest \ac{SRP} values. This procedure may continue for a maximum number of iterations, or until a minimum search cuboid is obtained. Note that this contraction procedure can also be applied to deterministic grids. In this case, the \ac{SRP} variant is referred to as \acf{CFRC} \cite{Do2007b}.

Although the aforementioned methods significantly accelerate the computation of \ac{SRP}, they provide no guarantees that the true source location will not be discarded, as they assume the SRP map to be a concave function with its maximum at the source location. The authors of \cite{Nunes2014a, Lima2015b, Lima2015c} propose a procedure which theoretically guarantees not to discard the point maximizing the \ac{SRP} function in anechoic conditions using the branch-and-bound iterative search method. The search starts by considering the entire search volume, typically the entire room, and subsequently divides it into smaller volumes using a branching function. Volumes are discarded through the aid of a bounding function similar to the bounds computed in \eqref{eq:v-srp}. 

Other iterative techniques used for \ac{SRP} include the Artificial Bee Colony \cite{Guo2015a}, Majorization-Minimization \cite{Scheibler2021, Togami2021} and Lagrange-Galerkin \cite{Liu2021} search methods. 

\subsection{Grids based on prior location estimates}
Alternatively, smaller grids can be built using lower-complexity, but less reliable source location estimators, such as those obtained using two-step methods. These candidates can then be more robustly selected and refined using \ac{SRP}. In \cite{Peterson2005c}, the grid is initialized using the positions associated with the signals' highest \ac{GCC-PHAT} peaks, which can be interpreted as estimates of the source's \ac{TDOA}. These estimates are used to triangulate candidate source positions using a least squares approach, which are then evaluated using \ac{SRP}. In \cite{Peterson2005c}, four peaks per pair were deemed to yield the best results.

As triangulation-based estimates are not robust to noise and reverberation, it is useful to include neighbouring points in the candidate grid, so as not to limit the performance of \ac{SRP}. This can be achieved by sampling points in the cuboid region containing these candidates \cite{Astapov2013c, Astapov2013e}. Similar approaches are proposed by \cite{Zhao2013c, Seifipour2013a, RanjkeshEskolaki2015, Awad-Alla2020a}. Grids based on prior location estimates were also explored on practical scenarios involving \acp{WASN} \cite{Astapov2013c, Astapov2013e}.

\subsection{Incorporation of prior scene information}

% Sensitivity
Another strategy for reducing the grid size exploits the property that spatial regions exhibit different levels of sensitivity depending on their position in relation to the microphone array \cite{Cho2009b, Yuan2012a, Yook2016a, Salvati2017a, Salvati2018b}. For instance, neighbouring candidate locations may have similar or identical sets of associated theoretical \acp{TDOA}, being therefore indistinguishable using \ac{SRP} \cite{Cho2009b, Yuan2012a, Yook2016a}. Those can therefore be replaced by their centroid without loss in performance \cite{Cho2009b, Yuan2012a, Yook2016a}.

A similar concept is proposed by \cite{Salvati2018b, Salvati2017a},
where a non-uniform, \textit{geometrically sampled grid}, is proposed. Based on their distance, each microphone pair within the system has a discrete set of integer \acp{TDOA}, in samples, each of which defines a hyperboloid in space. Candidate locations at the intersection of multiple hyperboloids have high definition, and can therefore be more reliably used for localization. Conversely, if the source is located within a low-definition region, more grid points are used to improve its localization performance.

% Prior microphone/source information grid stuff

Alternatively, information about the environment can be included as prior information to build smaller grids. For example, for specific microphone array geometries such as the T-shaped orthogonal array used by \cite{Cai2010b}, the 2D azimuth/elevation grid can be decomposed into two 1D grids, which can be independently maximized, significantly reducing the number of required \ac{SRP} evaluations. In \cite{DehghanFiroozabadi2010a}, a method combining \ac{SRP} for both \ac{DOA} estimation and \ac{PSSL} is proposed, and tested with a large aperture, L-shaped microphone array. \ac{SRP} is first used for estimating the source’s \ac{DOA} with respect to the array’s branches. This direction is used to create the initial grid of candidate locations, from which the \ac{SRC} variant of \ac{SRP} is employed for 3D localization. A similar two-step approach is employed in \cite{Zarghi2011a}, where distributed microphone arrays are used for \ac{DOA} estimation. The intersection of these directions is then used to estimate the source location. The computational complexity of SRP can also be reduced, for linear arrays, by combining array interpolation and polynomial root solving \cite{Johansson2005}. Alternatively, if possible source locations are known, such as seat locations in a conference room, a database of possible source locations along with their respective microphone array responses can be precomputed, thereby significantly reducing the grid size \cite{Zhuo2021a}.

% Pair selection
Complexity can also be reduced by reducing the number of pairwise maps computed. For instance, centralized microphone arrays of symmetrical geometries such as spherical or rectangular exhibit multiple pairs of microphones with parallel directions. Computation can be reduced at a negligeable loss in performance by only using one pair for each of those directions \cite{Grondin2022a}.  Conversely, microphone pair selection can also be applied to distributed microphone networks, where data transmission is a secondary constraint which should be minimized. If each device contains at least two microphones, the \ac{SRP} maps can be computed and transmitted independently for each device, an economic alternative to transmitting raw signals which was shown to incur only small losses in localization performance \cite{Cakmak2022a}.

Finally, the computation of the \ac{SRP} function can be avoided by only considering candidate positions with a high associated cross-correlation based on their theoretical \ac{TDOA} and \ac{GCC-PHAT} between microphone pairs \cite{Dmochowski2007e, Dmochowski2008b}. In practice, this can be achieved by creating a hash table for each microphone pair where each key-value pair represents a TDOA and its set of possible candidate positions. The keys (TDOAs) with a low associated GCC-PHAT can then be filtered out. Finally, the table is traversed, where the SRP values for the remaining sets of TDOAs associated with a candidate position are summed to create a global SRP map.

\subsection{Paralellization}

When the device computing \ac{SRP} supports parallel processing capabilities, such as multiple \acp{CPU}, multiple threads or one or more \acp{GPU},
the method can be sped up while using its original formulation, therefore guaranteeing its optimal performance. \ac{SRP} is highly parallelizable, as the evaluation of the SRP function for each candidate location is independent.

A \ac{CUDA} implementation of \ac{SRP} was first proposed in \cite{GonzagadaSilveiraJr2010a}, where the \ac{SRP} function for each candidate location was computed independently on each \ac{GPU} thread. In \cite{Minotto2013a}, a time-domain and a frequency-domain GPU implementation of \ac{SRP} using \ac{CUDA} were respectively compared with optimised \ac{CPU} counterparts. Results show the GPU implementations resulted respectively in speed improvements of 70 and 275 times. In \cite{Lee2016b}, the implementation provided by \cite{Minotto2013a} is optimised by maximising usage of the GPU’s internal memory in favour of the host’s memory, resulting in significant speed-up in comparison to \cite{Minotto2013a}.
In \cite{Belloch2013a}, an implementation of \ac{SRP} is proposed for three CUDA-enabled \ac{GPU} types.  In \cite{Belloch2015b, Belloch2015c}, a \ac{GPU} implementation of \ac{SRP} using NVIDIA’s Jetson chip, designed for low-power mobile computing, is evaluated for multiple grid resolutions. Conversely, in \cite{Belloch2015b}, a CUDA implementation of \ac{SRP} using multiple \acp{GPU} is presented.

In \cite{Lee2010a}, \ac{SRP}’s computation was vectorized using Intel’s \ac{IPP} software library, reducing \ac{CPU} load by a factor of two in comparison to a baseline scalar implementation. In \cite{Badia2019a}, an implementation of \ac{SRP} using OpenCL, an open-source parallel computing framework compatible with multiple processors including \acp{CPU}, \acp{GPU} and \acp{FPGA}, is presented. Experimental comparisons with device-specific implementations of \ac{SRP} reveal that the proposed implementation achieves similar performance. An efficient hardware implementation of \cite{Dietzen2021a} is presented in \cite{Yin2023a}.

\subsection{Other approaches}

In \cite{Grondin2019c}, an SRP method based on the singular value decomposition (SVD) is proposed. Based on \eqref{eq:srp_freq}, a matrix is defined mapping all frequency-domain GCCs to all candidate locations, whereof a low-rank approximation is obtained using the SVD. This low-rank approximation allows to first project frequency-domain GCCs onto a subspace with reduced dimensions and subsequently employing a k-d tree search scheme \cite{Bentley1975}, resulting in a lower computational cost at a similar localization performance to that obtained with the conventional SRP-PHAT. The performance of this method is increased in \cite{Grondin2019f}, where a spectral subraction procedure is applied to the correlation matrix.

It was shown in \cite{Dietzen2021a} that an SRP map can be efficiently approximated through interpolation while critically sampling the GCCs, based on Nyquist-Shannon sampling. Such approach is formulated while accounting for the physical bound over the range of possible TDOAs for a given microphone array, as well as the assumed GCC bandlimit. Simulation results indicate that the computational cost of the proposed interpolation-based approach for obtaining the approximated SRP map can be several orders of magnitude lower than the cost of computing the conventional SRP map, while the localization performance is maintained.

\section{Increasing robustness} \label{sec:robustness}

Although SRP has been shown to provide satisfactory performance in realistic scenarios \cite{Zhang2008a}, its performance is reduced in challenging scenarios including high reverberation and/or noise. Localization performance is often inversely related to the strategies presented in \autoref{sec:computational}, as fine grids provide better resolution. However, other techniques are required to remove artifacts caused by noise and reverberation from the \ac{SRP} maps.  

\subsection{Modified GCC-PHAT functions} \label{sec:rob:gcc}

% GCC-PHAT-beta
The quality of \ac{SRP} is dependent on the quality of the cross-correlation between microphone pairs. Most approaches employ \ac{GCC-PHAT} to obtain the correlation information, as it was shown to outperform temporal CC \cite{DiBiase2000a, Omologo1994a}. Nonetheless, modifications can be employed to improve \ac{GCC-PHAT} in challenging scenarios. One of such modification is $\text{GCC-PHAT}_\beta$, a parameterized version of $\text{GCC-PHAT}$ which was shown to improve localization performance, defined as \cite{Rabinkin1996, Donohue2007, Shen2009, Padois2019}
\small
\begin{equation} \label{eq:gcc-phat-beta}
    \text{GCC-PHAT}_\beta(f ; \bar{\mathbf{x}}_l, \bar{\mathbf{x}}_m) = \frac{\bar{\mathbf{x}}_l(f)\bar{\mathbf{x}}^{*}_m(f)}{\lvert \bar{\mathbf{x}}_l(f) \bar{\mathbf{x}}_m^*(f) \rvert^\beta + \gamma},
\end{equation}
\normalsize
where $\gamma$ provides numerical stability, and $\beta$ controls the relevance attributed to the signals' magnitudes. Note that conventional GCC-PHAT is achieved when $\beta = 1$, whereas conventional CC is obtained using $\beta = 0$. The experiments in \cite{Donohue2007} show that intermediary values of $\beta$ (e.g., $\beta=0.8$) improve localization of narrowband signals under the interference of directional noise sources at low \acp{SNR}. Although $\gamma$ is often set to a small value to prevent a null denominator, Shen et al. \cite{Shen2009} propose setting $\gamma$ to the minimum \textit{coherence} between the signal pair over all frequency bins. Coherence is here defined as the ratio between the signals' cross- and auto-spectral densities. In \cite{Ramamurthy2009a}, the authors perform an experimental analysis of the $\text{SRP-PHAT}_\beta$ method and they verify the simulation study in \cite{Donohue2007} which shows the acceptable range of values for the partial whitening parameter $\beta$ for a general signal to be between $0.65$ and $0.7$.  They also point out that the experiments exhibit more significant performance fluctuations for especially $\beta=1$ corresponding to the conventional PHAT method. This outcome supports the use of the partial whitening over the conventional PHAT.

% Other correlators
An alternative to PHAT filtering consists of using the kurtosis of the signal pair, motivated by the assumption that noise is frequently modelled as a Gaussian random process, which is theoretically eliminated in the kurtosis computation  \cite{Swartling2008a}. The GCCs can also be replaced by a sum of Gaussians centered at the former's most prominent peaks, thus producing a smoother \ac{SRP} map \cite{Cirillo2008a}. The effects of the phase transform can also be replaced by a linear predictor incorporating sparsity constraints \cite{He2018b}. The \ac{MCCC} function \cite{Benesty2004c} can also be employed \cite{Liu2022a}. Instead of providing a single correlation value for two signals and a delay $\tau$, \ac{MCCC} provides a correlation value for a vector of $M$ signals and a vector of delays $\pmb{\tau}$. The \acp{MCCC} can therefore be used to construct a beamformer which is applied as a preprocessing step before \ac{SRP} \cite{Liu2022a}. Finally, the CC between microphone pair signals can be computed using an eigenvalue decomposition of the cross-correlation matrix of the microphone signals. Instead of computing the CC between microphone signals, the correlation between corresponding eigenvectors can be used, ignoring directions related to noise and reverberation \cite{Wan2010a} and therefore improving the quality of the SRP map.

% Narrow-band stuff 
The \ac{GCC-PHAT} function of a broadband signal in an ideal, anechoic scenario is an impulse with its main peak occurring at the microphone pair's \ac{TDOA}. However, as the source signal becomes narrowband, the pair's \ac{GCC-PHAT} becomes a sinc function ($\text{sinc}(x) = \sin(x)/x$), i.e. a function exhibiting multiple ripples which translate into low-quality \ac{SRP} maps. In this case, the envelope of the \ac{GCC} function, obtained by extracting the magnitude of its analytic signal, can be applied instead to remove the aforementioned ripples.

% Broadband
In other broadband cases, some frequency bands may be more affected by noise than others. In those cases, it is advantageous to analyse the CCs in different frequency bands. This is done, for example in \cite{Cobos2020}, which proposes the creation of a GCC matrix, where columns represent frequency bands and rows represent time delays. The conventional \ac{GCC-PHAT} can be obtained from this matrix as long as the Constant Overlap-Add principle is satisfied when selecting the frequency band centres and widths. The authors show that degradations from noisy frequency bands can be reduced by applying SVD to obtain a low-order approximation of the GCC matrix, improving the robustness over the conventional GCC-PHAT.

% Outdoor, low-band
Many challenges also arise when applying SRP in large outdoor environments. Firstly, these environments suffer from intense low-frequency environmental noise, often requiring the signals to filtered before processing, thus creating a band-passed input signal which introduces challenges for the SRP method as described in \cite{Cobos2017d}. Secondly, the size of the search area may require very large grids, significantly increasing the method’s computational cost. Finally, factors such as changes in temperature, terrain, wind and position of the sensors make the propagation time model defined in \eqref{eq:propagation_time_distance} unreliable. The authors of \cite{Huang2021a} propose a modified GCC function based on Wavelet theory which takes the three aforementioned factors into account to improve the performance of \ac{SRP} in outdoor environments.

Finally, the \ac{GCC-PHAT} function can be substituted by a neural network \cite{Vera-Diaz2021a}, as will be discussed in \autoref{sec:nn}.

\subsection{Improving combination} \label{sec:rob:comb}
The formulation defined in \eqref{eq:srp_freq} combines pairwise and frequency-wise \ac{SRP} values through unweighted summation. A more general formulation of \ac{SRP}, which we denote Weighted SRP (W-SRP) can be written as
\small
\begin{equation} \label{eq:w-srp}
    \text{W-SRP}(\mathbf{u} \, ; \,\, \bar{\mathbf{X}}) = \bigcup_{(l, m) \in {M \choose 2}} \bigcap_{f \in \mathcal{F}} \frac{\text{SRP}_{lm}(\mathbf{u}, f \, ; \,\, 
    \bar{\mathbf{x}}_l, \bar{\mathbf{x}}_m)}{k_f k_{lm}},
\end{equation}
\normalsize
where $\bigcap$ represents the operation combining frequency information, $\bigcup$ represents the combination of pairwise information, and weighting factors $k_f$ and $k_{lm}$ respectively weight frequency and pairwise information. Besides classical summation, choices for the pairwise combinator $\bigcap$ are the product $\prod$ and the Hamacher t-norm, among others \cite{Pertilla2008}. 
Conventional SRP combines pairs through summation, meaning that pairwise \ac{SRP} maps combined in such manner will exhibit high values if any pair does so. Conversely, if multiplication is used, all pairwise maps must exhibit high values for the global \ac{SRP} to do so. In an extreme case, if any pairwise map is null, so will be the global \ac{SRP} map. The simulated experiments in \cite{Pertilla2008} show that combining pairwise \acp{SRP} through their product results in a significant increase in localisation performance over their sum, reducing the localization \acp{RMS} error by $45\%$.

% Pairwise weighting
The weights $k_{lm}$ can be computed on pairwise \ac{SRP} maps, for example, from a fractal theory standpoint, giving less importance to noisier, pairwise \acp{SRP} \cite{Wan2009}, or by measuring the noise of the \ac{GCC-PHAT} vector by computing the ratio between the \ac{GCC-PHAT}'s peak and its average \cite{Hummes2011a}. Note that microphone pair selection is also included in \eqref{eq:w-srp} for the special case $k_{lm} = \infty$.

Conversely, the frequency weight $k_f$ can be set as the maximum \ac{SRP} value across all pairs, therefore equalizing the contribution of each frequency bin to the global \ac{SRP}. This is shown to offer a similar effect to the PHAT weighting \cite{Salvati2014a}. Another approach estimates $k_f$ using neural networks \cite{Pertila2017a, Salvati2018c, Wang2019a, Wechsler2022a}, as will be discussed in \autoref{sec:nn}.

\subsection{Pre/Post-processing} \label{sec:rob:ppp}

% Pre-processing
Applying pre- or post-processing to the microphone signals in search of anomalies may improve \ac{SRP} maps. For example, a \ac{VAD} can be used to detect the presence of speech in a noisy environment, in order to prevent \ac{SRP} from unintentionally localizing noise sources \cite{Moragues2008a}, or to improve the localization of impulsive sources \cite{Machmer2009a}. A \ac{VAD} can also be used to discard directional noise sources \cite{Lim2015c}. \ac{SRP} maps can also be improved through the application of a Wiener filter \cite{Hummes2011a}.

\subsection{Neural approaches} \label{sec:nn}

As in many other tasks in acoustic signal processing, neural networks have also been applied for the task of \ac{SSL}, frequently obtaining state-of-the-art results in comparison to classical methods such as \ac{SRP} \cite{Grumiaux2021b}. However, \ac{SRP} still presents several advantages over classical neural network methods, which usually require matched training/testing microphone geometries. Furthermore, \ac{SRP} maps serve as an excellent input feature for neural networks. Finally, \ac{SRP}'s building blocks can be advantageously replaced by neural blocks, bridging the gap with neural methods' performance in challenging environments. The approaches below are related to the strategies mentioned in the above subsections.

% Neural-GCC
One of such blocks which can be improved is \ac{GCC-PHAT}. A deep neural block can be used to estimate an idealized \ac{GCC-PHAT} vector which removes peaks associated with reverberation and noise. A Deep-GCC function can be formulated in the time \cite{Vera-Diaz2021a, Qian2022} or frequency \cite{Yang2022b} domain. In the time domain, the Deep-GCC vector should exhibit a single peak at the source's true \ac{TDOA} $\tau_{lm}$, modeled as a Gaussian with standard deviation $\sigma_d$ as \cite{Vera-Diaz2021a, Qian2022}
\begin{equation} \label{eq:deep-gcc}
    \text{Deep-GCC}(\tau) = \exp \bigg(\frac{-|\tau - \tau_{lm}|^2}{2\sigma_d^2} \bigg).
\end{equation}
In practice, \eqref{eq:deep-gcc} serves as the target loss function for the network being trained. The choice of input feature and architecture for a Deep-GCC function may vary. In \cite{Vera-Diaz2021a, Qian2022}, \ac{GCC-PHAT} itself is chosen as the networks's input and a 1-D Convolutional autoencoder is selected as architecture. In \cite{Yang2022b}, the magnitude and phase spectrograms of both microphone signals are chosen as the input features, and a \ac{CRNN} is chosen as the neural architecture.

% Masking
Many approaches focus on using neural networks to estimate a weighting function, similarly to the signal processing based procedures described in \autoref{sec:rob:comb}. Most approaches focus on the frequency weights $k_f$, inspired by the task of speech enhancement, where neural time-frequency masks have attained significant success \cite{Pertila2017a}. For instance, a \ac{CNN} can be trained to estimate a time-frequency mask to reduce the interference of directional sources, using the output of a Wiener filter as its target function \cite{Pertila2017a}. Other targets can be used, such as the distance between the true and SRP-estimated locations for a single frequency band \cite{Salvati2018c, Wechsler2022a}.  Alternatively, the \ac{SNR} on each microphone can be used as a weight for each frequency band \cite{Wang2019a}. Similar approaches been also employed other machine learning methods, namely, \acp{SVM} and \acp{RBFN} \cite{Salvati2015, Salvati2016a, Salvati2016b}.

% post-processing/srp as a feature extractor
Another prominent manner of improving localization performance using \ac{SRP} uses the SRP maps as the input feature of a deep neural network. In this case, the neural network may have two goals: to enhance the maps produced by \ac{SRP} \cite{Diaz-Guerra2023, Diaz-Guerra2023a}, and/or to extract the source locations using the map \cite{Diaz-Guerra2018a, Diaz-Guerra2021a, zhao2021robust, Zhong2022a, Yin2023a, Grinstein2023b}, i.e., to improve the grid search/peak-picking function defined in \eqref{eq:vanilla-grid-search}. The networks differ in the architecture used, such as the \ac{MLP} \cite{Diaz-Guerra2018a, Grinstein2023b}, 3D \cite{Diaz-Guerra2021a, zhao2021robust, Yin2023a}, spherical \cite{Zhong2022a} and icosahedral \cite{Diaz-Guerra2023, Diaz-Guerra2023a} convolutions. %As neural networks are usually trained using backpropagation, all processing steps must be differentiable. Soft-argmax bla bla bla

\begin{figure}[h]
    \centering
        \includegraphics[scale=0.6]{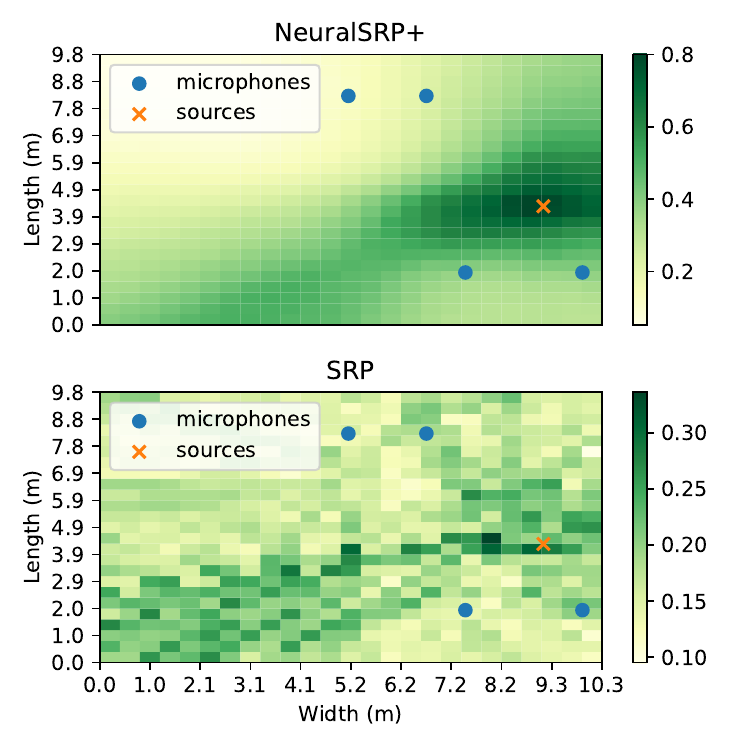}
    %\vspace{-0.8cm}
    \caption{Neural-SRP+ \cite{Grinstein2023c} and conventional SRP map in a highly reverberant room. The source position is shown with a cross and the microphone positions with circles. }
    \label{fig:neural-srp}
\end{figure}

% "Neural-SRP" approaches
Finally, other neural approaches simulate the pairwise processing used by \ac{SRP} for the task of source localization. The authors of \cite{Grinstein2023b} remark that the \ac{SRP} algorithm shares architectural similarities with the Relation Network, a type of \ac{GNN}. In the context of \ac{SRP}, a \textit{relation} between two microphones consists of the pairwise \ac{SRP} maps shared between them. All pairwise relations are then summed,  creating a global relationship between all microphones, which can be used to estimate the source locations. Neural-SRP approaches \cite{Grinstein2023c, Grinstein2023d} therefore replace \ac{SRP}'s function with a neural network, reducing the detrimental effects of noise and reverberation by including challenging scenarios during network training. An example of a map produced using a Neural-SRP method is shown in \autoref{fig:neural-srp}.

\subsection{Other approaches}

% Eric liked this one
\ac{SRP} maps can also be analysed by decomposing them using a set of idealized pairwise maps, computed using the theoretical \ac{TDOA} between the microphone pairs and the candidate locations. Instead of estimating the source location through peak-picking, the search can be done by matching the pairwise SRP maps with a subset of idealized maps according to a similarity metric \cite{Brutti2007a, Velasco2012}.

% Interpolation
When the distance between microphones in a centralized array is small, so is the range of possible \acp{TDOA} between the pair as expressed in \eqref{eq:max_tdoa}. It is therefore desirable to perform interpolation in the CC function to obtain sub-sample \ac{TDOA} resolution when using the temporal \ac{SRP} formulation. The work of \cite{Tervo2008a} evaluates the performance of \ac{SRP} for \ac{DOA} estimation from concert hall recordings using three different interpolation methods, namely, parabolic, exponential and Fourier. The study reports best performance using exponential peak interpolation.

The work in \cite{Asaei2009} presents a system where, before performing localization using SRP, a speaker verification step to remove unwanted speakers and noise is applied.

In \cite{Garcia-Barrios2023a}, the authors exploit spatial diversity in order to improve SRP’s performance in reverberant environments. Their simulation results show that large arrays are affected by the reverberation more than smaller ones and that having a smaller distance between microphone arrays results in more accurate localization. When the number of microphones in an array is increased the localization results are more robust as expected, but separating it into two array makes it even more favourable compared to merely increasing the number of microphones in a single array.

In \cite{DasH2016a}, a mel-frequency extraction technique is employed with \ac{SRP-PHAT} in order to obtain an enhancement of human speech and process it more robustly in a noisy environment. As a performance metric, peak SNR (PSNR) is used. The results show that utilizing \acp{MFCC} in conjunction with SRP-PHAT yields higher PSNR values compared to using only the SRP-PHAT, which results in a more accurate localization.

In \cite{Zhao2018a}, the authors compare the SRP-PHAT localization performances using a \ac{ULA} and a \ac{CPMA} interleaving two linear arrays with coprime dimensions. They show that a CPMA offers better localization results than a ULA with the same number of microphones. In another study \cite{Zhao2019a} by the same authors, a performance analysis of Semi-Coprime Microphone Arrays (SCPMAs) for localization using the SRP-PHAT algorithm is conducted. They evaluate the performance in terms of beam pattern, array gain and DOA estimation. The results on beam pattern an array gain suggest that the SCPMA outperforms the CPMA in reducing the peak side lobe level and minimising the total side lobe area. Moreover, it shows an enhanced ability to amplify the target signal while suppressing the noise. The results of DOA estimation in anechoic and low reverberant environments show that the SCPMA delivers accurate estimates which are on par with the estimates obtained from the full ULA. However, in highly reverberant conditions such as a $400$ ms reverberation time, side lobes in the beam pattern of the SCPMA result in less accurate estimates.

As discussed in \autoref{sec:near-far-field}, the range $\rho$ can only be accurately estimated when the source is located in the near-field with respect to the microphone array. The field type can be estimated by comparing the \ac{SRP} of two circular candidate grids at different distances, one in the far-field, the other in the near-field. The grid exhibiting the highest SRP value dictates the field regime. If near-field conditions are found, a second \ac{SRP} grid search can be applied for range estimation \cite{Zhao2013b}.

\section{Multi-source SRP approaches} \label{sec:multi-source}

We start this section by revisiting the problem statement described in \autoref{sec:base-model}. Instead of defining the target output of our system as a single source position vector $\mathbf{u}$, we extend it to be a matrix $\mathbf{U}$ of dimensions $3 \times N$, defined as
\begin{equation} \label{eq:matrix-U}
    \bf{U} = \begin{bmatrix}
        \mathbf{u}_1 & \mathbf{u}_2 & \hdots & \mathbf{u}_N
    \end{bmatrix},
\end{equation}
where $N$ is the number of active sources. Note that $N$ is usually unknown in practice, and must also be estimated on such cases. Updating the model for the signal received at each microphone is also required, as it becomes a weighted sum of all active sources. In the frequency domain, the received signal at microphone $m$ can be described as
\small
\begin{equation} \label{eq:received-multi-scalar}
    \bar{x}_m(t, f) = \sum_{n=1}^N s_n(t, f)a_{m}(\mathbf{u}_n, f)e^{-jf\tau_{m}(\mathbf{u}_n)} + \epsilon_m(t, f).
\end{equation}
\normalsize
Despite the modified signal model, the analysis of the CC function between two microphone signal frames $\mathbf{x}_l$ and $\mathbf{x}_m$ in the presence of $N$ simultaneous talkers usually presents one peak related to each source. Although this would allow the conventional SRP method to be used directly, the function may also exhibit `ghost peaks' related to the reflections caused by the room's surfaces, hindering the estimation procedure. Also, the relative amplitude of  peaks may vary considerably, especially in cases where the sources have different power levels, hindering the application of simple thresholding methods. Finally, the interfering sources reduce the correlation amplitudes at delays $\tau_{m}(\mathbf{u}_n)$ are reduced in comparison to the single source case, hindering the analysis of the SRP map.

Due to the aforementioned limitations of using the conventional SRP method for multi-source localization, different SRP-based alternatives have been proposed. These alternative methods, while presenting their own particularities in terms of implementation, target scenario and performance, are categorized in the following subsections based on their core modification when compared to the conventional SRP method.

\subsection{Modified SRP computation}
\label{sec:multisource:preproc}
Different strategies have been proposed where the computation of the SRP map is modified in order to allow for better localization of multiple sources.
For instance, in \cite{Donohue2007}, an alternative to the conventional PHAT-weighting function is proposed, aiming to achieve flexibility in combining different narrowband components. Simulation results, obtained for both single and multi-source cases, indicate that the use of the modified PHAT-weighting function can improve localization performance for both narrowband and broadband signals. 

In \cite{Padois2016}, similarly to the efforts aimed at achieving an improved combination of pair-wise information for increasing localization robustness  outlined in \autoref{sec:rob:comb}, the use of harmonic and geometric means of the GCC functions over all available microphone pairs was explored to build an acoustic map. When compared to the conventional summation of pair-wise functions, as previously expressed in \eqref{eq:srp}, results show that the use of geometric and harmonic means contributes to removing undesired sidelobes and improving source level estimation.

\subsection{Source cancellation}
\label{sec:multisource:postproc:sourcecancel}

\begin{figure}
\centering
    \includegraphics[width=0.9\columnwidth]{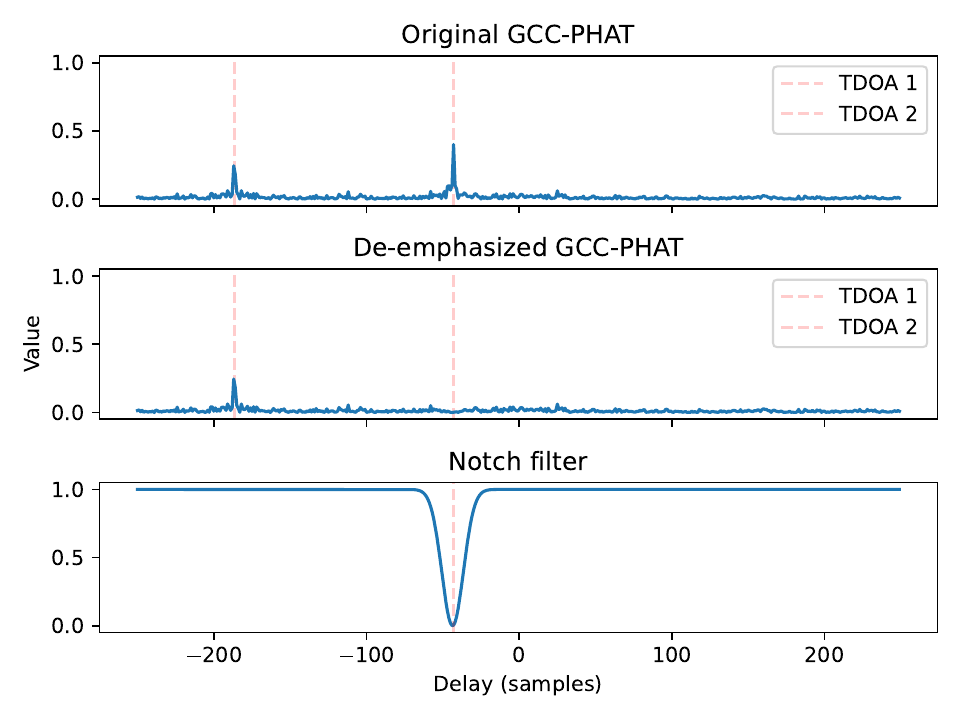}
  \caption{Representation of the de-emphasis procedure described by Brutti et al. \cite{Brutti2010a}.}
   \label{fig:de-emphasis}
\end{figure}

Many multi-source, SRP-based methods rely on schemes that reduce the influence of a previously located and dominant source on newly computed SRP maps. For instance, in \cite{Brutti2010a}, the localization of two sources is performed in a two-step manner. First, the position of the source with the highest correlation peak is estimated as in the conventional \ac{SRP} method. To estimate the second source, the first source is de-emphasized from the CC function through the use of a TDOA-domain notch filter. This process is illustrated in \autoref{fig:de-emphasis}. Although this approach can be further applied for the localization of three sources, the authors state that the noise in the correlation function with three sources would be prohibitive, and that tracking approaches should be applied instead.

The removal of a previously located source's contribution from an SRP map can also be achieved through the projection of the observed GCCs onto a subspace that is orthogonal to the source position, as described in \cite{diaz2020source}. Results obtained with both simulated and experimental data indicate that such an approach can outperform the de-emphasis method from \cite{Brutti2010a}, especially in cases of sources with different power levels. Moreover, the use of a truncated formulation of the proposed source cancellation scheme allows for a reduction in computational cost while performing comparably to \cite{Brutti2010a}, without requiring parameter tuning associated to the TDOA-domain notch filter design.

Subspace processing for source cancellation within an SRP-based framework has also been proposed in \cite{Grondin2019e}, where the SVD-PHAT method \cite{Grondin2019c} is  extended to address the case of multiple sources. Thererin, the contribution of a previously located source (obtained by means of a k-d tree search) is removed from the observed projections of the GCCs onto a reduced-dimensional subspace. The proposed multi-source SVD-PHAT approach was compared to a source cancellation scheme, similar to the de-emphasis method from \cite{Brutti2010a}, where a source's contribution is removed from the observed GCCs and a new SRP-PHAT map is computed for locating the next source. Simulation results indicate that the multi-source SVD-PHAT can outperform the successive recomputation of the SRP-PHAT map.

As an alternative to employing a source cancellation procedure to the observed GCCs, the spatial gradient SRP-PHAT method proposed in \cite{Taghizadeh2011a} involves successively removing the influence of the current, most dominant source directly in the observed SRP map by means of a negative spatial gradient function. Experimental results for two-speaker scenarios show that the spatial gradient SRP-PHAT can be an effective localization method in scenarios with a diffuse noise field. 

In \cite{Oualil2012a}, an approximate analytical formulation of an SRP map using a Gaussian Mixture Model is proposed, such that probability density functions can be used to estimate the location of multiple sources while removing their corresponding contributions from the probabilistic SRP map. Experimental results with scenarios involving up to three speakers indicate that while this approach can effectively locate multiple sources, its performance degrades when sources differ greatly in power.

\subsection{Grid refinement}
\label{sec:multisource:postproc:gridref}

In \cite{Oualil2013a}, grid refinement is indirectly used to localize multiple sources by identifying different zones of interest, defined in terms of TDOA intervals, as those where acoustic sources are dominant in terms of a cumulative SRP function. In this way, a conventional grid search step for source localization can be performed over a reduced search space with the desired spatial resolution. The localization of multiple sources can then be achieved by iteratively removing the influence of the dominant sources \cite{Oualil2012a}. Experimental results show that such approach can improve localization performance in multi-source scenarios at a lower computational cost than the authors' previous work \cite{Oualil2012a}.

In \cite{Coteli2018}, a hierarchical search-grid refinement method is proposed, where a probability measure of a sound source's presence in different regions, formulated as a spatially averaged SRP map, is used to identify the limited set of steering directions for which the search grid resolution can then be improved for localizing multiple sources. This approach is shown to lower the computational cost while performing similarly to the conventional SRP method that employs the highest resolution level over the entire search space.

\subsection{Clustering and distance analysis}
\label{sec:multisource:postproc:cluster}

Another concept often exploited in multi-source localization methods relates to data clustering and analyzing distances between multiple source location estimates. For instance, the sources' preliminary location estimates can be obtained through the conventional SRP method. Then, spatial clustering can be employed to track the estimated locations of multiple sources over different time frames \cite{Segura2008a}. Alternatively, a narrowband SRP formulation can be employed to obtain location estimates per frequency bin and time frame, while Gaussian mixture modeling can then be used to cluster the location estimates \cite{Madhu2008a}. Furthermore, both the location and activity of multiple sources can be tracked \cite{Madhu2008a}.

In \cite{Do2008}, source location estimates are obtained by using SRP-PHAT combined with agglomerative spatial clustering and SRC (cf. \autoref{sec:iterative}). Experimental results show that the localization performance of the proposed approach degrades when the peaks to be identified have widely different amplitudes or are closely located in the CC function. Accordingly, the proposed approach is further extended in \cite{Do2010a}, by replacing the agglomerative clustering step from \cite{Do2008} with Gaussian mixture modeling of the observed SRP map, or by identifying the peaks in the SRP map while assuming a minimal distance between  sources. The performance limitations first demonstrated in \cite{Do2008} are also addressed in \cite{Cai2010c}, where the localization of multiple speech sources is achieved by computing subband SRP maps, estimating the dominant source's position for each subband, and employing agglomerative clustering across all subbands to obtain the final set of source location estimates. In \cite{Khanal2015a}, a method named Multi-Stage Rejection Sampling (MSRS) is proposed, which involves spatially clustering probability density points, derived as a function of the observed SRP-PHAT map, to identify regions of interest. Then, volume contraction is used in the identified regions for localizing multiple sources.

In \cite{Boora2020}, a three-step framework is proposed for multiple source localization. It relies on: step 1) partitioning the search region into cubic volumes, clustering such volumes and, based on equivalent TDOA bounds; step 2) computing a delay density map to find in which clusters it is more likely to have a sound source; step 3) further analyzing the chosen clusters with conventional SRP to obtain the final source location estimates. 

Finally, the approach proposed in \cite{Firoozabadi2022} for a specific microphone setup of central and lateral microphone arrays, involves finding the intersection between the source positions estimated with the central array's SRP map and the ones estimated with the lateral arrays through an adaptive subband generalized eigenvalue decomposition (GEVD) scheme, in order to obtain the final 3D location estimates of multiple sources. Simulation results with up to three speakers demonstrate that the proposed method outperforms other state-of-the-art methods under varying levels of noise and reverberation.

\subsection{Sparsity assumptions}
\label{sec:multisource:postproc:sparsity}
Sparsity-based modifications to the conventional SRP method have also been proposed for the task of multiple source localization. For instance, by assuming a limited number of active sources with respect to the search grid of candidate locations, localization can be performed by employing a sparse-regularized generative model that fits the observed SRP map, combined with a subspace filtering step that compensates for what is not directly accounted for by the fitted model \cite{Velasco2012}. Experimental results show that although the use of this approach can outperform the conventional SRP-PHAT in the multi-source scenarios tested, its overall performance highly depends on the  choice of the hyperparameters used in the proposed problem formulation.

Alternatively, in \cite{Tengan2023}, it is shown that group sparsity can be exploited when modeling an observed broadband SRP map as a linear function of power spectral densities (PSDs), related to an overcomplete set of candidate locations. Hence, multi-source localization can be achieved by solving a group-sparse optimization problem and identifying peaks in the estimated PSDs. Simulation results obtained for two-speaker scenarios show that the proposed method performs better than or similar to the conventional SRP-PHAT method for varying levels of noise and reverberation, while overall outperforming the frequency-domain
sparse iterative covariance-based estimation (SPICE) \cite{Stoica2011,Park2018} method.

In \cite{Pavlidi2013}, the authors exploit time-frequency sparsity, by assuming that only one speech source is dominant in a given time-frequency bin, i.e., they are assumed to be W-disjoint \cite{Rickard2002}. By analyzing each frequency bin and performing single-source localization, histograms with all individual DOA estimates can be generated and used in a matching-pursuit-based step to perform multiple source localization. Simulation and experimental results indicate that this approach can outperform other state-of-the-art multi-source localization methods, at a lower computational cost. The sparsity of speech signals in the time-frequency domain is similarly exploited in \cite{Hadad2018}, where a weighted, wideband histogram of source locations is computed based on narrowband DOA estimates, obtained with SRP-PHAT applied to different frequencies and observation frames. The weighted histogram is then used to perform multiple source localization through peak detection, and simulation results indicate the advantage of the proposed method when compared to the wideband SRP-PHAT for two-speaker scenarios in reverberant environments.

In \cite{Padois2017}, sparse modeling of the GCCs observed by a microphone array is employed in the task of localizing sound sources and their corresponding acoustic reflections. A linear inverse problem is proposed to be solved, with its formulation depending on a time-domain propagation matrix. The authors present two implementations of the proposed method, with the first based on orthogonal match pursuit (OMP) \cite{Pati1993a}, and the second on the truncated Newton interior-point method \cite{Kim2007a}. It is demonstrated through an experimental study that the use of sparsity constraints in the solution of the proposed linear inverse problem contributes to better location estimates when compared to the direct use of a time-domain SRP map. The choice of propagation matrix used for formulating the linear inverse problem presented in \cite{Padois2017} was further investigated in \cite{Chu2020}, where the influence of the temporal width threshold, associated to the determination of propagation matrix coefficients, is demonstrated. Additionally, when assuming the GCC coefficients to be PHAT-weighted, an alternative formulation of the propagation matrix circumventing such temporal width threshold is proposed, with experimental results indicating the advantage of using such alternative formulation in terms of computational time.

Finally, in \cite{Thakallapalli2021}, an SRP-based method is proposed for simultaneous multiple source localization that employs \ac{NMF} \cite{Lee2000} to decompose the time-frequency signal into a weighted sum of broadband atoms, which are time-frame-dependent and correspond to different groupings of frequency bands related to distinct sources. This method, named SRP-NMF, attempts to combine the advantages of both narrowband and broadband approaches that exploit sparsity in their corresponding domains, and experimental results indicate it performs better than or similarly to state-of-the-art methods based on fully broadband or narrowband signal formulations.

\section{Practical considerations} \label{sec:applications}

\subsection{Applications}
\ac{SSL} is a foundational task which has been applied in many domains, having been used as an input feature for speech enhancement/beamforming tasks \cite{Bai2008a, Levi2010b, Traa2016a}, voice activity detection \cite{Blauth2012a, Schwartz2018a}, speaker diarization \cite{Traa2015a, Nikunen2018, Cabanas-Molero2018a, Kang2020a, Gburrek2022b}, sound source separation \cite{Dam2016a, Dam2017a, Wu2021a} and array calibration \cite{Hennecke2009a}. Furthermore, \ac{SRP}'s localization performance can be improved by combining it with other sensors, such as LIDAR \cite{Even2012a, Even2013} or multi-sensor devices \cite{Seewald2014a}. \ac{SRP} has been used on the multiple practical scenarios described below.

% Surveillance and defense
Although \ac{SRP} can be used to localize any type of sound source, many applications focus on a specific sound event. A prominent application is that of surveillance and defence. \ac{SRP} can be used to localize irregular \acp{UAV} activity \cite{Sedunov2018a, Harvey2019}, as well as using an \ac{UAV} with an embedded microphone array to localize sources of interest itself \cite{Strauss2018a, Tengan2024PhD}. Other applications in security include intrusion detection \cite{Zieger2009a, Kim2020}, and gunshot localization \cite{Lopez-Morillas2016a, Park2021a}.

% Outdoor/large
Another category of interest is that of scene understanding in large and/or outdoor environments, such as the detection of indoor and outdoor sources of noise pollution \cite{Chiariotti2019, Royvaran2021a, Tiete2014} and the detection of underground seismic events \cite{Nie2019a}. \ac{SRP} was also applied for commercial and environmental purposes, such as the localization of sound-emitting fish using an underwater hydrophone \cite{DeVille2019a}, and to detect faulty equipment within electrical power stations \cite{Chen2020b}. Furthermore, with the increased interest in smart and self-driving vehicles sensors, localization of horns and crashes using \ac{SRP} \cite{Do2007c, Shon2012a} can also be performed, or localizing talkers inside the vehicle itself \cite{Swerdlow2008}.

% Indoor/medical/corporate
Turning to indoor environments, \ac{SRP} can be applied to the medical domain, being used to localize and analyze footsteps with the goal of early detection of dementia \cite{VanDenBroeck2013}, as well as for fall detection of elderly people \cite{Li2012a}. \ac{SRP} can also be used to improve human-robot interactions \cite{Wang2004, Lebarbenchon2018, Gamboa-Montero2022a}, as well as for camera steering corporate meetings \cite{Marti2011b} and smart rooms \cite{Johansson2002a, Butko2011a}. \ac{SRP} was also applied to a helmet-mounted microphone array \cite{Zhang2014}, which can be used for increasing acoustic awareness on industrial sites, for example.

\subsection{Tracking moving sources} \label{sec:tracking}
Although a source may remain mostly stationary in many scenarios such as conference calls, the same cannot be said for many situations in surveillance, robotics and healthcare. It is therefore reasonable to reformulate the source position 
$\mathbf{u}$ to be time-dependent, i.e., $\mathbf{u}(t)$. The task of estimating a source's position at multiple time instants is hereafter referred to as tracking.

A straightforward way to achieve tracking using conventional \ac{SRP} is to compute an SRP map and estimate the source position independently for successive frames at times $t_{i - 1}$ and $t_{i}$. This estimate can be often improved through the incorporation of a state-space dynamic model as well as previous estimates $\{ \hat{\mathbf{u}}(t_{i - 1}) \;\; \hat{\mathbf{u}}(t_{i - 2}) \ldots \}$. Such a state-space model provides source tracking by introducing dynamic constraints into the source localization procedure, modeling for instance the speed of the source. This allows for smoother position estimates to be produced and for unreliable observations, such as those caused by reverberation and noise, to be properly identified and handled.

The most common approaches for source tracking using \ac{SRP} are Kalman filters \cite{Abad2006a, Astapov2014a, Segura2008a, Grondin2019d}, particle filters \cite{Ward2003a, Valin2006b, Habib2010b, Fallon2012a, Wu2013b, Astapov2014a, Wu2016, Wang2018f} and deep neural networks \cite{Diaz-Guerra2021a, Yang2022b, Zhong2022a, Diaz-Guerra2023}. Unlike in neural methods, the state-space model is explicitly defined in Kalman and particle filters. 

Particle filters are frequently preferred over Kalman filters due to their simpler formulation and ability to model non-linear systems. Particle filters model the source location with the help of $Q$ candidate positions known as particles, each having an associated likelihood or weight $\pi_q$, $q=1,\,\hdots,\,Q$. The estimated source location is obtained as a weighted sum of the particles, where the weights are their respective likelihood. At each iteration, the particles are updated according to a given kinematic model. Optionally, a resampling process may be also applied to reduce the variance of the particles.

The movement of a source at consecutive time steps is commonly modeled using Langevin dynamics \cite{Ward2003a, Habib2010b, Fallon2010a, Fallon2012a, Wu2013b, Wu2016}, which assume that the source moves independently in each direction. The relationship between the source's position at times $t_{i - 1}$ and $t_i$ is equal to \cite{Fallon2010a}
\begin{equation} \label{eq:langevin-position}
    \mathbf{u}(t_i) = \mathbf{u}(t_{i - 1}) + \dot{\mathbf{u}}(t_i) \Delta t,
\end{equation}
where $\Delta t = (t_{i - 1} - t_i)/f_s$, and $\dot{\mathbf{u}}(t_i)$ is the source's velocity, modelled as
\begin{equation} \label{eq:langevin-velocity}
    \dot{\mathbf{u}}(t_i) = a^{(1)} \dot{\mathbf{u}}(t_{i-1}) + b^{(1)} F^{(1)}.
\end{equation}
In \eqref{eq:langevin-velocity}, $F^{(1)} = {N}(0, 1)$, $a^{(1)} = e^{-\alpha^{(1)} \Delta t}$ and $b^{(1)} = \beta^{(1)} \sqrt{1 - a^{(1)}}$ are known as the damping and excitation parameters, respectively responsible for controlling the inertia and innovation of the movement in each direction. $\alpha^{(1)}$, $\beta^{(1)}$ are hyperparameters, to be chosen or tuned.

\subsection{Directional sources and microphones} \label{sec:directivity}

The SRP signal model can be modified for the case where sources and/or microphones exhibit directional acoustic behaviour, that is, the amplitudes of the microphone signals are dependent on the orientation of microphones and sources. The directivity profile for microphone $m$ is defined as a function $0 < d^{(1)}_m(\theta_m) \leq 1$,  where $\theta_m$ is an angle. An analogous function can also be defined for the source's directivity $d^{(2)}(\theta_s)$. Finally, we define angles $\theta_1$, $\theta_2$, $\theta_3$ and $\theta_4$ as the angles of departure, the source direction, the angle of arrival and microphone direction respectively. The attenuation term defined in \eqref{eq:prop-anechoic} can then be specified as \cite{Mungamuru2004a}
\begin{equation} \label{eq:attenuation}
    a_m = d^{(1)}_m(\theta_1 - \theta_2) d^{(2)}(\theta_3 - \theta_4)\frac{k_d}{\lVert \mathbf{u} - \mathbf{v}_m \rVert},
\end{equation}
where $\frac{k_d}{\lVert \mathbf{u} - \mathbf{v}_m \rVert}$ represents the attenuation caused by propagation, which generally follows an inverse law. In practice, this attenuation can be incorporated into \ac{SRP} by including the source's candidate orientation as another search dimension \cite{Mungamuru2004a}. Note that the gains between microphones must be assumed to be calibrated, and that the source and microphone directivity patterns, as well as the microphone orientations,  must be known or assumed. Microphone directivity can also be exploited to reduce the number of microphone pairs and region size used for \ac{SRP} \cite{Grondin2019d, Grondin2022a}.

When operating with distributed microphone arrays, source directivity can be estimated in two steps, firstly by estimating the source position, followed by the creation of a spherical grid around the source. The point with the highest \ac{SRP} value on the sphere is selected to represent the source's orientation \cite{Brutti2005a, Brutti2006a}. A similar approach is applied in \cite{Nakadai2005, Abad2007, Togami2010a}, which assumes that the arrays directly facing the speaker will exhibit an SRP map with a sharp peak. The sharpness is measured using the map's kurtosis, which is then used to estimate the talkers orientation. If the microphone gains are calibrated, the \ac{GCC-PHAT}'s peak values can be used for comparison instead of the kurtosis \cite{Segura2012a}.

\subsection{Comparing SRP to other approaches}

% ML-SSL description
In \cite{Zhang2008a}, a theoretical comparison between the \ac{SRP} and the \ac{ML-SSL} method is made. The functioning of the \ac{ML-SSL} method is similar to \ac{SRP}, as the source location is also estimated as the maximum argument of a likelihood function. However, the \ac{ML-SSL} method explicitly models the noise received at each microphone as well as its correlation with other microphones. Although this can be advantageous, allowing microphone signals exhibiting large noise to be ignored, it requires noise statistics to be assumed or measured. The aforementioned paper starts by defining the \ac{ML-SSL} signal model similarly to \eqref{eq:prop-reverb} directly in the frequency domain. The authors assume that reverberation is independent across microphone signals, and that microphones boast a high \ac{SNR}. The paper shows that, under these assumptions, the \ac{ML-SSL} method becomes independent of noise and reverberation, and equal to the \ac{SRP} formulation, and uses this proof to justify why \ac{SRP} works well under low-noise, reverberant environments.

In \cite{Silverman2005a}, the authors conducted a performance analysis of several GCC-PHAT-based algorithms for a large-aperture microphone array. They presented a real-time source localization algorithm based on TDOAs derived from a phase transform applied to the generalized cross-power spectrum. The algorithm is then enhanced by preprocessing the data using local beamformers. A comparison was made by testing these two algorithms and the SRP-PHAT in an environment in which the microphone signal-to-reverberation ratio was in the range [$-2$ dB, $-12$ dB] and the signal-to-background noise ratio with flat frequency weighting in the band 80 Hz to 10 kHz was in the range [$-6$ dB, $-16$ dB]. They found that SRP-PHAT provides reliable location estimates under adverse conditions but has a larger computational cost.

In \cite{Nguyen2011a}, the authors developed two GCC methods based on time delay estimation using classical CC and smoothed coherence transform algorithms. They analyzed the performance of the aforementioned GCC-based algorithms for multi-source, point-based localization by comparing them with the existing FASTTDE, GCC-PHAT and FAST SRP-PHAT \cite{lathoud-rr-06-26}. The methods were evaluated in terms of 1) localization accuracy, 2) detection accuracy, and 3) computational cost. The localization accuracy of the FAST SRP-PHAT was much higher than that of the other three methods. In terms of detection, the other three methods exhibited higher localization performance. Finally, it was shown that SRP-PHAT had a higher computational cost than other methods. 

In \cite{Johansson2004b}, a comparison between the ROOT-MUSIC algorithm and SRP-PHAT is made. The root mean square error (RMSE) is used as a performance metric and the results show that even though ROOT-MUSIC is more computationally efficient, SRP-PHAT exhibits superior performance in challenging conditions, such as environments with reverberation and low SNR.

In \cite{Dmochowski2007d, Dmochowski2010b}, the authors evaluate the performance of broadband spatio-spectral estimators, including SRP and two-step localization methods. They perform an eigenanalysis of the parameterized spatial correlation matrix and show that the attenuation can be estimated from this matrix. They propose a DOA estimator based on MCCC and show that this method yields a higher resolution than the conventional SRP. DOA estimation performance is similar in anechoic environments and environments with reverberation time of $300$ ms, however it is worse in environments with reverberation time of $600$ ms compared to SRP.

In \cite{Hafezi2016b}, the authors evaluate the performance of a multiple source localization method based on \ac{AIV} using spherical microphone arrays. Their simulations comprising various angular separation and reverberation times show that AIV has an average accuracy between $5$ and $10$ degrees for sources with angular separation of $30$ degrees or more and it performs better than the methods using Pseudo-Intensity Vectors and SRP. Another finding is that a plane wave decomposition-based SRP method cannot localize all sources if the number of sources is three or more and if they are separated less than $45$ degrees. %revise

In \cite{Blauth2012a}, SRP-PHAT is used along with Hidden Markov Models (HMMs) and face tracking for voice activity detection and localization. Results show that using HMMs along with SRP-PHAT increases the accuracy and utilising face tracking in addition yields even better results.

 In \cite{Peterson2005b}, the authors compared two algorithms \cite{Zotkin2004b}, \cite{Peterson2005c}, which are also  explained in \autoref{sec:computational}, in both simulated and real-world scenarios in which speakers were recorded by eight microphones spread out on the wall and ceiling, and concluded that hybrid localization is more robust than hierarchical localization and is computationally faster. Especially when the reverberation time is greater than 300 ms, the localization error in hierarchical localization increases at a much higher rate.

In \cite{Aarabi2003a}, a general framework for the integration of microphone signals for \ac{SSL} is presented. A \ac{SOF}, which is the mean square difference between the \ac{SLF} and the true probability of an object, is used as an indication of the accuracy of the map. SRP-PHAT is a special case of this method in which the SLF from each array is integrated without taking the SOFs into account.

\subsection{Analyses of SRP} \label{sec:analysis}

The authors of \cite{Velasco2016a} propose an analytical model based on sound propagation and its interaction with the environment that predicts SRP maps in both anechoic and non-anechoic conditions, and under both far- and near-field assumptions. They investigate how and to what extent the signal bandwidth, array topology, room geometry and spectral content of the signal affect SRP maps. The findings show that SRP functions depend on the array topology, room geometry and signal bandwidth but not on the spectral content of the signal. They validate their model by comparing it with the true SRP maps. 

In \cite{Swartling2011a}, the authors investigated the geometrical sensor calibration errors in a \ac{ULA} used in far-field human speech source localization. They observed that the errors in configuration of the endpoint sensors result in larger localization errors compared with same configuration errors of the inner sensors. In addition, they show that the localization errors increase when the total configuration error is above a threshold related to the propagation distance and the system's sampling rate.

In \cite{Nie2022a}, the authors proposed an SRP constraint to suppress local extrema. They weighted the SRP function using a coherence factor, determined by observing the signs of the GCCs between all possible microphone pairs. If the sign was the same for all microphone pairs, this indicated a high coherence and the coherence factor is $1$. If half of them were negative and half were positive, then were deemed as incoherent, and assigned a coherence factor of $0$. The method was shown to operate without loss of localization accuracy with respect to conventional \ac{SRP}.

% Contribution
\section{X-SRP} \label{sec:xsrp}

\begin{figure*}
    \begin{center}
        \includegraphics[width=0.9\textwidth]{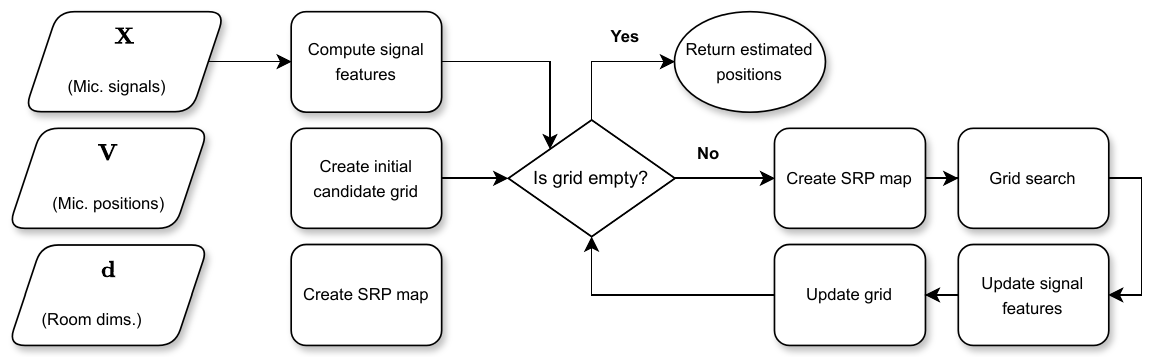}
    \end{center}
    \caption{Flowchart of the generalized SRP algorithm. Parallelograms represent input data, rectangles represent functions, diamonds represent decisions and ellipses represent terminal states.}
    \label{fig:xsrp-block}
\end{figure*}

In this section, we describe the \ac{SRP} method from an algorithmic perspective, with the goal of unifying the previously described extensions of the method within a common framework. The main functionality described in \autoref{sec:base-model} is revisited by substituting specific functions with generic ones, which we shall refer to as \textit{modules}. For example, the CC function used in the classical \ac{SRP} is substituted by a module called \text{compute\_signal\_features}, which can be instantiated as the temporal CC, \ac{GCC-PHAT}, or a neural-based feature as in \cite{Pertila2017a, Yang2022b}. 

This modular perspective allows for \ac{SRP} papers to be grouped in an alternative way to the task-oriented manner used in the previous sections. Conversely, the categorization presented here groups the works by their implementation details, facilitating their combination and comparison.

To facilitate the reproduction of SRP variants and explore novel variants, we release the eXtensible-SRP, or X-SRP Python library, which provides a modular implementation of \ac{SRP} following \autoref{alg:srp}, which is also shown as a flow diagram in \autoref{fig:xsrp-block}. We include multiple modules within \ac{XSRP}, which allow for selected variants to be implemented. We refer the reader to the project's repository \footnote{\url{https://github.com/egrinstein/xsrp}} for further documentation on the library.

\begin{algorithm} \caption{X-SRP}
\label{alg:srp}
\begin{algorithmic}[1]
\Function{x-srp}{$\mathbf{X}, \mathbf{V}, \mathbf{d}=\text{null}$}
\State $ \hat{\mathcal{U}} \gets \emptyset $
\State $ \mathbf{G} \gets \text{create\_initial\_candidate\_grid($\mathbf{d}$)}$
\State $ \mathbf{C} \gets \text{compute\_signal\_features($\mathbf{X}$)}$

\While {$\mathbf{G} \neq \emptyset$}
    \State $\mathbf{S} \gets \text{create\_srp\_map}(\mathbf{V}, \mathbf{G}, \mathbf{C})$
    \State $\hat{\mathcal{U}} = \text{grid\_search}(\mathbf{G}, \mathbf{S}, \hat{\mathcal{U}})$
    \State $\mathcal{C} = \text{update\_signal\_features}(\mathcal{C}, \hat{\mathcal{U}}, \mathbf{V})$
    \State $\mathbf{G} = \text{update\_grid}(\hat{\mathcal{U}}, \mathbf{d})$
    
\EndFor
\State \Return {$\hat{\mathcal{U}}$}
\EndFunction
\end{algorithmic}
\end{algorithm}

\Autoref{alg:srp} accepts three input parameters: A matrix $\mathbf{X}$ of microphone signal frames, a matrix of microphone positions $\mathbf{V}$ and a vector $\mathbf{d}$ containing the room dimensions. It is made optional as it is only necessary for \ac{SSL}, not for \ac{DOA} estimation. Note that configuration parameters such as the sampling rate $f_s$ are omitted for the sake of conciseness. 

The first line initializes the estimated source coordinates $\hat{\mathcal{U}}$ as an empty set. $\hat{\mathcal{U}}$ is a set of points and not a single point to accommodate multi-source localization approaches.

Then, an initial grid of candidate positions $\mathbf{G}$ is created using the \textbf{create\_initial\_candidate\_grid} module. In most \ac{SRP} variants, this will be the only grid created. However, in iterative approaches such as \cite{Nunes2014a, Do2007c} as well as multi-source approaches \cite{Brutti2010a}, this function only provides an initial grid $\mathbf{G}$, which is further updated as part of their grid search or refinement procedure. Typically, the grid created is a 2D or 3D Cartesian grid for \ac{SSL}, or a polar or spherical grid for \ac{DOA} estimation. In the latter case, the room dimensions $\mathbf{d}$ are not used, as the grid is produced with respect to the microphone array's centre. This grid can typically be computed as a pre-processing step if the microphone and room geometries are known beforehand.

The \textbf{compute\_signal\_features} module computes $\mathbf{C}$, which can be the CC function between microphone pairs, their \ac{GCC-PHAT}, or neural-based features \cite{Pertila2017a, Yang2022b}. 
% In conventional \ac{SRP}, each spatial value is mapped to the CC delay closest to its actual \ac{TDOA}. In Volumetric SRP approaches \cite{Cobos2011a, Lima2015b}, each grid point is mapped to several cross-correlation values pertaining to the point's volume of influence. In \cite{Dmochowski2007e, Dmochowski2008b}, the map is inverted: each delay is mapped to several candidate positions. Finally, in interpolation approaches \cite{Tervo2008a}, each position is mapped to their actual \acp{TDOA}.

Line (5) begins a loop, which represents the grid search procedure. For most approaches, this loop will only execute once, and will only execute multiple times in approaches such as \cite{Nunes2014a, Do2007c}. We define as the loop's stopping criterion the candidate grid $\mathbf{G}$ being empty, symbolizing that end of the grid search.

The module \textbf{create\_srp\_map} computes the \ac{SRP} map $\mathbf{S}$, which assigns a likelihood value to each grid point in $\mathbf{G}$, using the microphone positions $\mathbf{V}$ and the temporal features $\mathbf{C}$. 

Then, the \textbf{\text{grid\_search}} module searches for the grid points in $\mathbf{G}$ that maximize the \ac{SRP} map $\mathbf{S}$ to estimate the source coordinates $\hat{\mathcal{U}}$, as well as a new grid of candidate locations $\mathbf{G}$. When localizing a single source, \text{grid\_search} returns $\hat{\mathcal{P}} = \{\argmax_{\mathbf{G}} \mathbf{S}\}$, and an empty grid, i.e., $\mathbf{G} = \emptyset$.

The \textbf{\text{update\_signal\_features}} module is used to alter the signal features $\mathbf{C}$. This is mainly used in iterative and multi-source approaches such as the source de-emphasis procedure \cite{Brutti2010a} (cf. \autoref{fig:de-emphasis}). Finally, the \textbf{\text{update\_grid}} module may be used to generate a new grid based on the current source estimates $\hat{\mathcal{U}}$. An example of variant using this grid is the SRC approach \cite{Do2007c}.

\section{Conclusion} \label{sec:conclusion}

% \begin{figure}
%     \begin{center}
%         \includegraphics{images/6.pdf}%[width=0.4\textwidth]{images/6.pdf}
%     \end{center}
%     \caption{(a) Mean localization error for DI-NNs and baselines on different datasets. (b) Normalized histogram comparison between the DI-NN and the CRNN baseline on the recorded dataset. (c) Cumulative version of (b).}
%     \label{fig:errors}
% \end{figure}

In this paper, we showed that the \ac{SRP} method remains an important localization method, and is still under continuous improvement. We hope that the detailed description of the conventional \ac{SRP} method, followed by a presentation of the combination of the literature into multiple categories has allowed the reader to learn or increase their knowledge on \ac{SRP}. Finally, we hope that the alternative division of \ac{SRP} into functional blocks will allow for the method to be further expanded.

Future research directions on \ac{SRP} include further improvement of neural methods, by allowing an arbitrary number of sources to be concurrently detected, inclusion of prior information such as noise statistics as a secondary network input, or architectural modifications, for example. Signal processing-based \ac{SRP} modifications can also be improved by exploring other types of multi-source and tracking strategies, as well as alternative strategies for combining pairwise and frequency-wise information.

\backmatter

% \bmhead{Acknowledgements}

% Acknowledgements are not compulsory. Where included they should be brief. Grant or contribution numbers may be acknowledged.

% Please refer to Journal-level guidance for any specific requirements.

\section*{Declarations}

\subsection{Availability of data and materials}
Code used for generating figures and simulations used on this paper is available in \href{GitHub}{https://github.com/egrinstein/xsrp}

\subsection{Competing interests}

N/A

\subsection{Funding}

The research leading to these results has received funding from the European Union's Horizon 2020 research and innovation programme under the Marie Skłodowska-Curie grant agreement No. 956962 and from the European Research Council under the European Union's Horizon 2020 research and innovation program / ERC Consolidator Grant: SONORA (no. 773268). This paper reflects only the authors' views and the Union is not liable for any use that may be made of the contained information.

\subsection{Authors' contributions}

E.G.: manuscript writing, simulations and coding. E.T.: manuscript writing and simulations. B.Ç., T.D., L.N., T.vW. M.B. and P.A.N.: Manuscript writing.  

\subsection{Acknowledgements}
N/A

% \begin{appendices}

% \section{Section title of first appendix}\label{secA1}

% An appendix contains supplementary information that is not an essential part of the text itself but which may be helpful in providing a more comprehensive understanding of the research problem or it is information that is too cumbersome to be included in the body of the paper.

% \end{appendices}

\bibliography{sapstrings, SRP}

\end{document}